\documentclass{article}

\usepackage[main, final]{neurips_2026}

\usepackage[utf8]{inputenc}
\usepackage[T1]{fontenc}
\usepackage{hyperref}
\usepackage{url}
\usepackage{booktabs}
\usepackage{amsfonts}
\usepackage{nicefrac}
\usepackage{microtype}
\usepackage{xcolor}
\usepackage{wrapfig}
\usepackage{colortbl}
\usepackage{float}
\usepackage{multirow}
\usepackage{caption}
\usepackage{graphicx}
\usepackage{amsmath, bm}
\usepackage{lipsum}
\usepackage{mathrsfs}
\usepackage{fancyhdr}
\usepackage[ruled]{algorithm2e}
\usepackage{enumitem}

\usepackage{algorithmic}
\usepackage{setspace}
\usepackage{amssymb}
\usepackage{makecell}
\usepackage{subcaption}

\title{Backdoor Attacks under  Lossy Compression: From Failure to Reactivation and Adaptation}

\author{%
  \textbf{Qian Li$^{1,2}$}
  ~~\qquad\textbf{Yunuo Chen$^{1}$}
  ~~\qquad\textbf{Yuntian Chen$^{2}$}\thanks{Corresponding Author} \\
  $^{1}$Shanghai Jiao Tong University~~
  $^{2}$Eastern Institute of Technology \\
}

\begin{document}

\maketitle

\begin{abstract}
\label{abstract__}
Real-world backdoor attacks often require poisoned datasets to be stored and transmitted before they are used to compromise deep learning systems. In the era of big data, however, the inevitable use of lossy compression poses a fundamental challenge to invisible backdoor attacks. We observe that triggers embedded in RGB images can become ineffective once the images are lossily compressed into binary bitstreams, such as JPEG files, for storage and transmission. Consequently, poisoned data may lose their malicious functionality after compression, causing backdoor injection to fail.
Prior compression-based attacks typically exploit compression artifacts as certain triggers to distinguish poisoned RGB samples from uncompressed benign ones, rather than addressing whether malicious information can survive a shared lossy storage-and-transmission pipeline.
In this paper, we highlight the necessity of explicitly accounting for lossy compression in backdoor attacks. This requires attackers to ensure that transmitted binary bitstreams preserve malicious trigger information, such that effective triggers can be induced after decompression.
Building on the region-of-interest (ROI) coding mechanism in image compression, we propose two poisoning strategies tailored to inevitable lossy compression. First, we introduce \textbf{Universal Attack Reactivation}, a general method that uses sample-specific ROI masks to reactivate trigger information in  bitstreams for learned image compression (LIC). Second, we present \textbf{Compression-Adapted Attack}, a new attack strategy that employs customized ROI masks to encode trigger information into  bitstreams and applies to both traditional codecs and LIC. Extensive experiments demonstrate the effectiveness of both strategies.
\end{abstract}

\vspace{-5pt}
\section{Introduction}\label{intro}
\vspace{-5pt}

In recent years, the widespread deployment of deep learning systems has exposed them to a broad range of security threats~\cite{carlini2017towards,duan2024conditional,LQ1,costales2020live,doan2020februus,doan2021lira,cohen2019certified,gu2017badnets,cox2007digital,ganju2018property,LQ2,liu2018trojaning,gao2019strip}. Among them, data poisoning-based backdoor attacks pose a particularly severe risk. In such attacks, adversaries inject malicious ``triggers'' into a subset of training data and release the poisoned dataset to downstream users~\cite{li2020invisible,lin2020composite,nguyen2020input,saha2020hidden,salem2022dynamic}. Models trained on such datasets behave normally on benign inputs, but exhibit attacker-controlled predictions once the trigger is present. Due to their stealthiness and strong controllability, backdoor attacks have become a major threat to the security and reliability of machine learning systems.

Most existing backdoor attacks are designed and evaluated in an idealized RGB-domain setting, where poisoned images are assumed to be directly used for model training after trigger injection. However, this assumption often deviates from real-world data pipelines. In practical scenarios, poisoned datasets typically need to be stored, transmitted, downloaded, and decoded before they are used to compromise deep learning systems. With the rapid growth of data volume and increasing demand for efficient storage and network transmission, lossy image compression has become increasingly inevitable~\cite{jamil2023learning,hu2021learning,yang2023lossy}. This introduces a fundamental challenge to invisible backdoor attacks: triggers embedded in RGB images may not survive the conversion from RGB data to compressed binary bitstreams and back to decompressed RGB images.

Formally, many invisible attacks construct poisoned samples by directly modifying RGB images, e.g., ${\bm x}_p={\bm x}+{\bm \delta}\in\mathbb{R}^{3\times H\times W}$. Yet, once ${\bm x}_p$ is lossily compressed into a binary bitstream $\hat{\bm y}\in\{0,1\}^n$, such as a JPEG file, there is no guarantee that the malicious trigger information is preserved in $\hat{\bm y}$. After decompression, the recovered poisoned image $\hat{\bm x}_p\in\mathbb{R}^{3\times H\times W}$ may no longer contain an effective trigger. Since lossy compression inevitably introduces distortion, i.e., $\hat{\bm x}\neq{\bm x}$, it can attenuate or destroy trigger signals. As a result, poisoned data that are effective in the raw RGB domain may lose their malicious functionality after storage and transmission, causing backdoor injection to fail. This phenomenon affects a broad range of existing invisible attacks, including both invisible-trigger methods and clean-label attacks~\cite{bppattack,Ftrojan,wanet,issba,sleeper,narcissus}.

\begin{wrapfigure}{r}{0.52\linewidth}
  \vspace{-15pt}
  \centering
  \includegraphics[width=\linewidth]{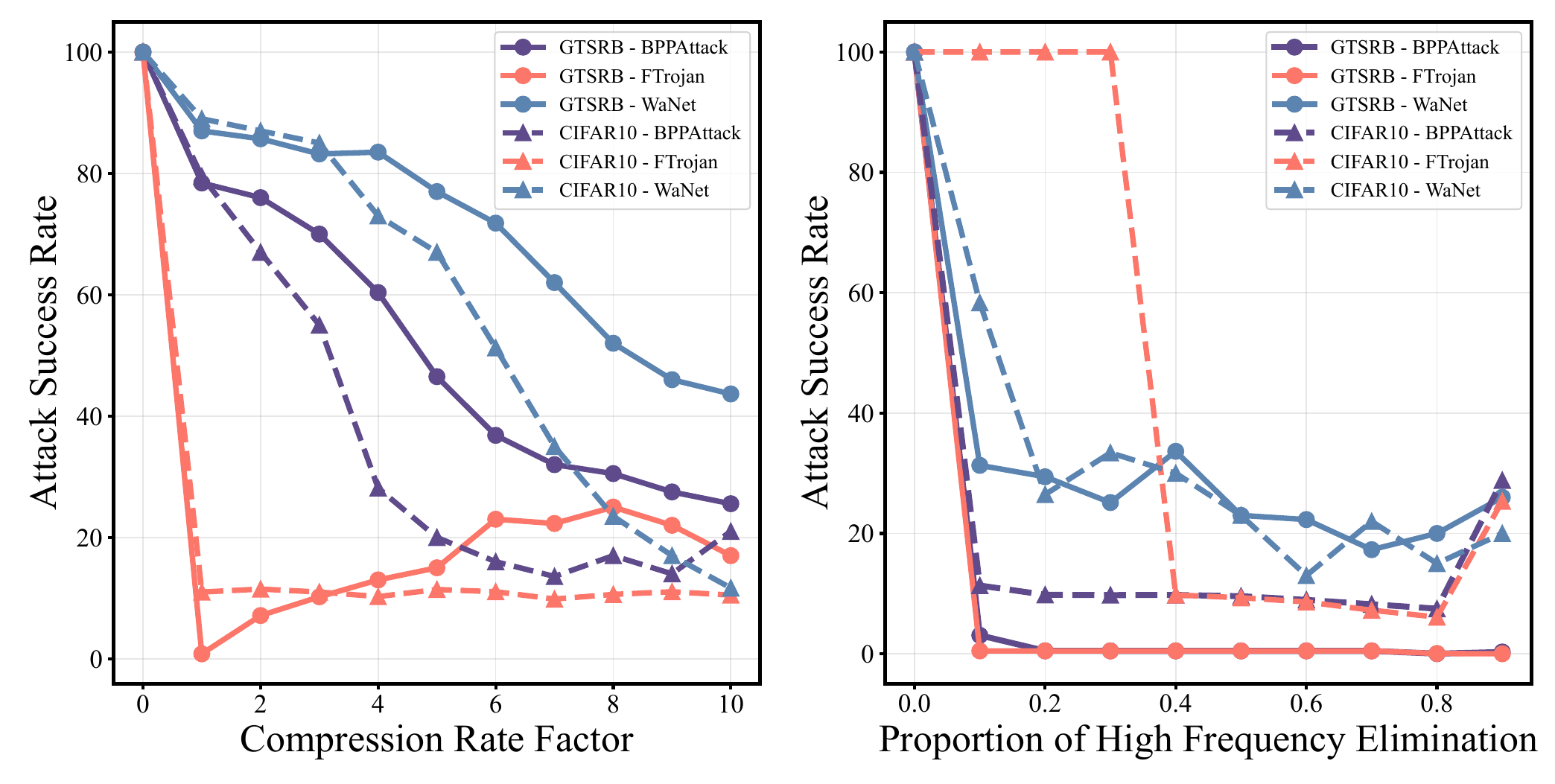}
  \vspace{-15pt}
  \caption{Ineffectiveness of backdoor attacks under lossy compression.
    \textbf{Left:} Higher compression rates introduce stronger distortion and substantially hinder backdoor injection, indicating that lossy compression can severely damage invisible triggers.
    \textbf{Right:} Removing high-frequency components from poisoned test samples via Fourier Transform significantly reduces the ASR, suggesting that high-frequency information is critical for many invisible triggers.}
  \label{fig:combined}
  \vspace{-8pt}
\end{wrapfigure}

We observe that this failure is closely related to the frequency characteristics of invisible triggers. Inspired by prior studies on frequency-domain backdoor behavior~\cite{zeng2021rethinking,iclrrethinking}, we find that many invisible triggers rely on subtle high-frequency components to achieve stealthiness. These components are visually unobtrusive, but often play a crucial role in triggering the backdoor. As shown in Fig.~\ref{fig:combined}, stronger compression substantially reduces the attack success rate, and explicitly removing high-frequency components from poisoned test samples also leads to a significant ASR drop. These observations suggest that invisible triggers achieve stealth by hiding in compression-sensitive high-frequency information, while lossy compression tends to suppress or distort exactly such information. Therefore, ignoring the compression stage can lead to overly optimistic evaluations of backdoor attacks and render poisoned datasets ineffective in real-world storage-and-transmission pipelines.

It is worth distinguishing our setting from prior compression-based backdoor attacks. \textit{Existing compression-based methods \cite{yang2023everyone, duan2024conditional} typically remain trigger-centric}: they exploit compression artifacts as triggers to distinguish poisoned RGB samples from \textit{uncompressed benign samples}. In contrast, they do not explicitly address whether malicious information can survive a shared lossy storage-and-transmission pipeline, where \textit{both benign and poisoned data are compressed into binary bitstreams}. Our goal is therefore not merely to use compression artifacts as another form of trigger. Instead, we study backdoor attacks under inevitable lossy compression, where the attacker must ensure that the transmitted bitstream recovers or induces malicious trigger information after decompression.

To address this challenge, we explicitly incorporate lossy compression into the design of backdoor attacks. Rather than evaluating the toxicity of raw RGB poisoned images alone, attackers must ensure that both the compressed bitstream $\hat{\bm y}$ and the decompressed image $\hat{\bm x}_p$ carry effective malicious information. We achieve this by leveraging the Region of Interest (ROI) coding mechanism in image compression~\cite{roi5,song2021variable,roi1,roi2,roi3}. ROI coding allows codecs to assign different bitrates to different image regions, thereby controlling the spatial distribution of compression distortion. This provides a natural interface for manipulating how trigger-related information is preserved or encoded during compression, without modifying the standard decompression process used by victims.

Following this principle, we propose two ROI-based poisoning strategies tailored to inevitable lossy compression from failure to reactivation and adaptation. The first, \textbf{Universal Attack Reactivation}, aims to reactivate existing invisible attacks that become ineffective after compression by preserving trigger-relevant information during learned image compression (LIC)~\cite{xie2021enhanced,zou2022devil,li2022hybrid,unoc,chen2026nefic}. The second, \textbf{Compression-Adapted Attack} (CAA), goes one step further by designing triggers that are inherently adapted to the compression process, using customized ROI masks to encode malicious information into compressed bitstreams. Together, these two strategies cover complementary paths: reactivating fragile RGB-domain triggers after compression, and constructing compression-adapted triggers that remain effective through lossy storage and transmission.

Our contributions are summarized as follows:
\begin{itemize}
  \item[$\bullet$] We reveal that previously effective invisible triggers can become ineffective after lossy compression. This finding highlights the necessity of explicitly accounting for inevitable storage-and-transmission compression in data poisoning-based backdoor attacks.

  \item[$\bullet$] We propose \textbf{Universal Attack Reactivation}, a general ROI-based method for LICs that reactivates invisible triggers degraded by lossy compression. Extensive experiments show that it significantly improves the attack success rates of prior invisible attacks under compression.

  \item[$\bullet$] We propose \textbf{Compression-Adapted Attack} (CAA), a new invisible backdoor attack for both LICs and traditional codecs. CAA encodes trigger information through customized ROI masks and achieves strong attack performance and stealthiness under  lossy compression.
\end{itemize}

\vspace{-5pt}
\section{Related Work}\label{RW}
\vspace{-5pt}
\subsection{Invisible Backdoor Attack}
\vspace{-3pt}
Gu \textit{et al.} first introduced a backdoor attack by their BadNets proposal~\cite{gu2017badnets}, which has sparked a lot of research interest on backdoor attack~\cite{chen2017targeted, liu2018trojaning, liu2020reflection, cheng2021deep, huynh2024combat_add1, jiang2023color_add2}. Recently, researchers have increasingly focused on the invisibility of backdoor triggers to develop more powerful attacks. ISSBA~\cite{li2021invisible} embedded class-specific strings into images using  steganography techniques. WaNet~\cite{wanet} applied image warping as triggers, making the modifications hard to notice.  BppAttack~\cite{bppattack} took advantage of vulnerabilities in the human visual system, reducing visible changes through image quantization and improving trigger accuracy with contrastive learning. FTrojan~\cite{Ftrojan} used perturbations in the frequency domain, spreading subtle pixel-wise changes throughout the image. Additionally, Rao \textit{et al.}~\cite{iclrrethinking} proposed a backdoor attack method based on  semantic information in the frequency domain.
Clean-label attack~\cite{sleeper,narcissus,saha,turner} is a special type of invisible backdoor attack that implants a backdoor without altering sample labels, enabling stealthier data poisoning. Its triggers are typically tiny, high-frequency adversarial perturbations and are therefore especially vulnerable to distortion from lossy compression.
However, existing invisible backdoors overlook lossy compression effects, which remove high-frequency trigger components and can severely degrade attack performance.

\vspace{-5pt}
\subsection{Lossy Compression and Compression-Aware Backdoor Attacks}
\vspace{-5pt}
Lossy image compression reduces storage costs by discarding perceptually redundant information, making it increasingly important as image data volumes grow. Traditional codecs such as JPEG \cite{ROIJPEG}, JPEG2000\cite{jpeg2000}, and BPG \cite{ROIBPG} rely on hand-crafted transforms and entropy coding.  More recently, learned image compression (LIC) has achieved stronger rate-distortion performance with end-to-end  neural networks and  a variational autoencoder framework ~\cite{balle2016end, balle2018variational, chen2022two, zhu2022transformer, liu2023learned, zhang2023neural, iccv25HPCM, zeng2025mambaic, chen2026adaptive, he2021checkerboard, minnen2020channel, qian2022entroformer, jiang2023mlic, li2023frequency, chen2026contentaware,li2025differentiable}, driving rapid progress in image compression.

Several recent studies have also explored the connection between lossy compression and backdoor attacks. Existing compression-based attacks~\cite{yang2023everyone,duan2024conditional} typically remain trigger-centric: they exploit compression artifacts as triggers to distinguish poisoned RGB samples from \textit{uncompressed benign samples}. Therefore, their effectiveness largely relies on a distributional gap between compressed poisoned samples and uncompressed benign samples in the RGB domain. In the inevitable compression setting considered in this paper, however, both benign and poisoned data are stored and transmitted through a shared lossy compression pipeline. As a result, the compression-artifact triggers used by these methods  no longer provide a reliable discriminative signal, and can become ineffective similarly to conventional invisible triggers after compression and decompression (discussed in Sec. \ref{exps}).

Another related line of work studies compression-resistant backdoor attacks~\cite{xue2023compression-resist, yu2023backdoor, yu2024robust, wang5268955gaba}. While such methods improve robustness against compression, they typically require the attacker to participate in or control the training process of a specific classification model, for example by modifying the training objective. This differs from the dataset-only poisoning setting considered in this paper, where the attacker only releases a poisoned dataset and the victim trains a model using an arbitrary standard training pipeline. Our work focuses on this more practical setting and explicitly asks whether malicious information can survive storage and transmission as compressed bitstreams and be recovered or induced after decompression. Thus, lossy compression is not merely treated as a trigger source, but as an unavoidable communication channel in real-world data poisoning pipelines.

\begin{figure*}[t]
\centering
\includegraphics[width=0.91\textwidth]{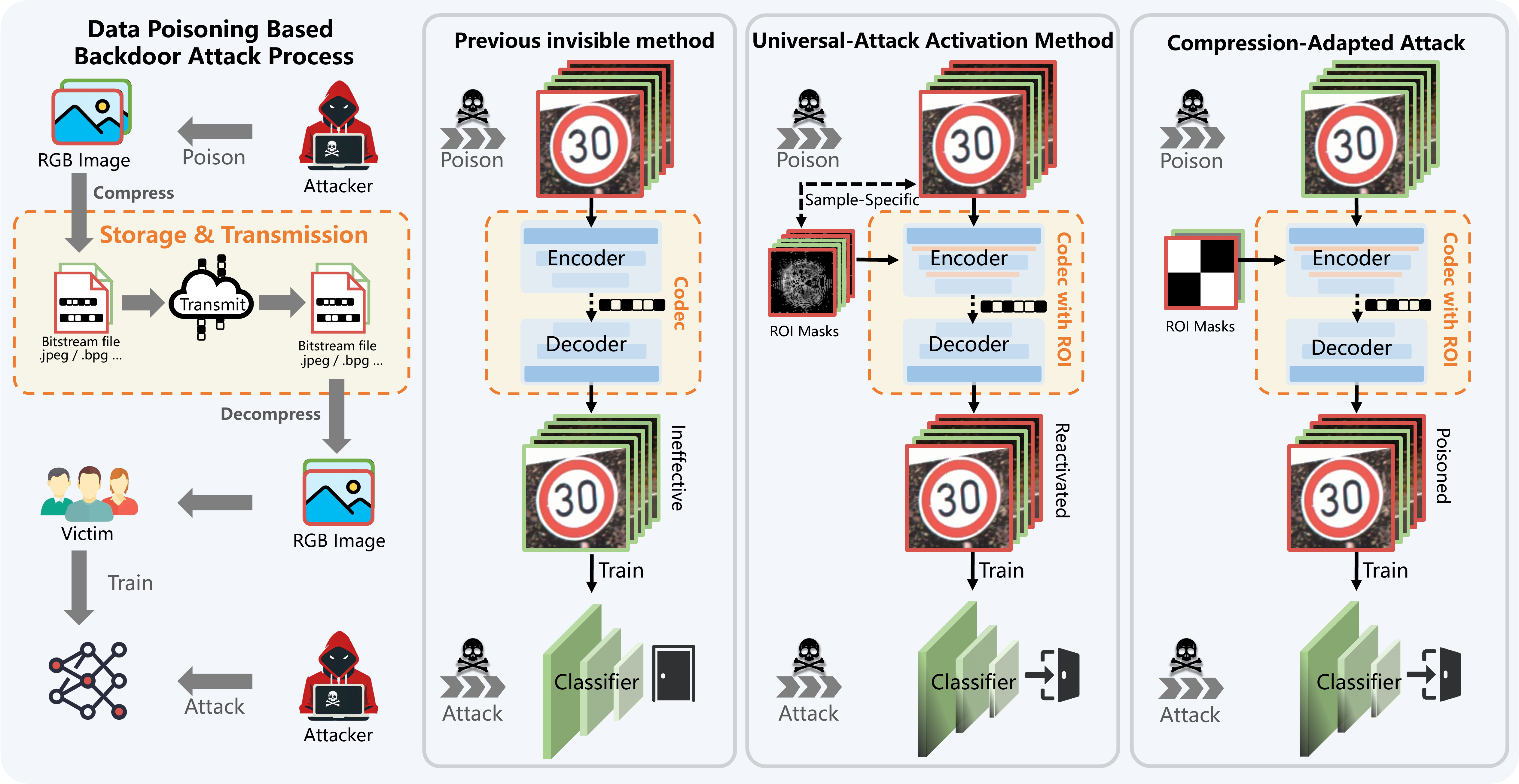}
\vspace{-5pt}
\caption{\textbf{Overview of backdoor attacks under  lossy compression.}
Red borders denote poisoned samples, and green borders denote benign samples.
\textbf{Column 1} shows the real-world dataset-only poisoning pipeline, where data are stored and transmitted as compressed bitstreams before victim-side training.
\textbf{Column 2} illustrates that previous invisible attacks may fail after lossy compression because trigger information is destroyed.
\textbf{Columns 3-4} present our two ROI-based strategies: \textbf{Universal Attack Reactivation}, which reactivates ineffective invisible attacks by preserving trigger-relevant information, and \textbf{Compression-Adapted Attack}, which encodes malicious information into bitstreams to construct compression-adapted triggers for both traditional codecs and LICs.}
\label{Fintro}
\vspace{-15pt}
\end{figure*}

\vspace{-10pt}
\section{Backdoor Attacks under Lossy Compression}
\label{method}
\vspace{-8pt}

\subsection{Threat Model and Problem Setup under Lossy Compression}
\vspace{-8pt}

Fig.~\ref{Fintro} illustrates the dataset-only poisoning pipeline considered in this paper.
Unlike the idealized RGB-domain setting, poisoned and benign images are first stored and transmitted as bitstreams before being used for victim-side training.
Given an image ${\bm x}\in\mathbb{R}^{3\times H\times W}$, a lossy codec encodes it into a binary bitstream $\bm y$, quantizes it to $\hat{\bm y}$, and later decodes it back into a reconstructed RGB image $\hat{\bm x}$.:
\begin{equation}
\begin{aligned}
{\rm Compression:}\quad \hat{\bm y}&={\rm Q}({\rm Encoder}({\bm x})),\quad
{\rm Decompression:}\quad \hat{\bm x}&={\rm Decoder}({\rm DeQ}(\hat{\bm y})).
\end{aligned}
\label{EnDe}
\end{equation}
where ${\rm Q}$ and ${\rm DeQ}$ denote quantization and dequantization, respectively.
This compression--decompression process is central to our setting.
Previous invisible attacks usually construct poisoned samples in the RGB domain, e.g., ${\bm x}_p={\bm x}+{\bm \delta}$, but do not ensure that the trigger information survives the mapping ${\bm x}_p\rightarrow\hat{\bm y}_p\rightarrow\hat{\bm x}_p$.
As lossy compression introduces distortion, the trigger embedded in ${\bm x}_p$ may be attenuated or removed from the bitstream $\hat{\bm y}_p$, making the decompressed sample $\hat{\bm x}_p$ ineffective for backdoor injection.
Therefore, attacks under inevitable lossy compression must explicitly ensure that malicious information is preserved, recovered, or induced through the compressed bitstream.
We consider the following dataset-only poisoning setting:

\textbf{1. Targets.}
The attacker aims to release a poisoned dataset whose decompressed samples can successfully implant a backdoor into a victim-trained model, while maintaining normal accuracy on decompressed benign samples.
We consider storage and transmission with publicly available codec implementations or specifications, so the attack is evaluated under realistic  compression pipelines.

\textbf{2. Capabilities.}
The attacker can poison data before storage and transmission and can control the sender-side encoding process, including configuring or modifying the encoder using public codec specifications and parameters.
The output, however, must be a standard-compliant bitstream that can be decoded by the victim's unchanged decoder.

\textbf{3. Constraints.}
The attacker cannot modify the victim-side decoder, training algorithm, or model architecture, nor can they require the victim to use a specially designed decoder.
All poisoned files must be valid compressed bitstreams with normal decoding behavior, normal bitrate and file size, and visually normal decompressed images.

\vspace{-8pt}
\subsection{ROI-Supported Compression}\label{F-t}
\vspace{-8pt}

To manipulate how information is preserved during compression without modifying the decoder, we leverage Region of Interest (ROI) coding, a common functionality in image compression~\cite{ROIJPEG,ROIBPG,roi5,roi6,song2021variable,roi1}.
An ROI-supported encoder takes an additional mask ${\bm M}=[m_{ij}]\in\mathbb{R}^{H\times W}$ as input:
\begin{align}
    {\rm Compression:}\quad \hat{\bm y}={\rm Q}({\rm Encoder}({\bm x},{\bm M})).\label{En2}
\end{align}
The ROI mask is used only as an encoder-side control signal. Each mask value $m_{ij}$ controls the bitrate allocation weight at spatial location $(i,j)$.
A larger weight allocates more bits to the corresponding region and thus reduces compression distortion, while a smaller weight allows stronger distortion.
Therefore, ROI coding provides a natural interface for controlling which image information is preserved in the transmitted bitstream.

This interface is well suited to our goal.
Instead of introducing additional backdoor modules or requiring a modified decoder, ROI masks allow the attacker to guide the standard compression process itself.
We use this mechanism to design two complementary attack routes under lossy compression.
The first route, introduced in Sec.~\ref{RIBA}, targets LIC models with pixel-level ROI support and reactivates existing invisible triggers by preserving trigger-relevant information.
The second route, introduced in Sec.~\ref{CABA}, targets both traditional codecs and LICs with region-level ROI support and constructs compression-adapted triggers by encoding malicious information through customized ROI masks.

For LICs, ROI support can be integrated through spatially adaptive modules such as Spatial Feature Transform (SFT)~\cite{wang2018recovering_sft, song2021variable}.
To satisfy our constraints, such modules are used only on the encoder side, while the decoder remains standard and unchanged.
For traditional codecs, we directly rely on existing ROI-supported implementations~\cite{ROIJPEG,ROIBPG,jpeg2000}.

\vspace{-8pt}
\subsection{Universal Attack Reactivation (UAR): Reactivating Fragile Triggers}\label{RIBA}
\vspace{-8pt}
Existing invisible backdoor attacks provide diverse trigger designs, but mostly  become ineffective after  compression as the trigger information is degraded under compression.
Instead of redesigning these attacks, our first route aims to \emph{reactivate} them by guiding the compression process to preserve trigger-relevant information.
This strategy is particularly suitable for LICs with pixel-level ROI support, where fine-grained bitrate allocation can be controlled through sample-specific ROI masks.

Given a clean image ${\bm x}$ and its poisoned counterpart ${\bm x}_p$ generated by an existing attack, a direct way to identify trigger-relevant regions is to measure their residuals as ROI masks:
\begin{equation}
{\bm M}_{\rm res}={\rm Norm}\big({\rm Mean}|{\bm x}-{\bm x}_p|\big),
\end{equation}
where ${\rm Mean}(\cdot)$ averages over RGB channels, and
${\rm Norm}({\bm z})=\frac{{\bm z}-{\rm min}({\bm z})}{{\rm max}({\bm z})-{\rm min}({\bm z})}$ normalizes the mask values.
This mask allocates more bits to regions with larger RGB-domain residuals, which is especially useful for additive or localized invisible triggers.

However, residual magnitude alone may not fully capture the information that is most vulnerable to compression.
As discussed in Sec.~\ref{intro} and Fig.~\ref{fig:combined}, many invisible triggers rely on subtle high-frequency components, which are crucial for attack activation but are also heavily distorted by lossy compression.
Motivated by this observation, we further design a frequency-based ROI mask to preserve compression-sensitive trigger components.
Specifically, we extract high-frequency information from ${\bm x}_p$ and map it back to the image domain:
\begin{equation}
{\bm M}_{\rm freq}=
{\rm Norm}\big({\rm Mean}|\mathcal{F}^{-1}(h_t(\mathcal{F}({\bm x}_p)))|\big),
\end{equation}
where $\mathcal{F}$ denotes the Fast Fourier Transform and $h_t$ is a high-frequency truncation function with threshold $t$.
Compared with ${\bm M}_{\rm res}$, ${\bm M}_{\rm freq}$ focuses on preserving the high-frequency trigger information that is most likely to be suppressed during compression.

With either ${\bm M}_{\rm res}$ or ${\bm M}_{\rm freq}$, the decompressed poisoned sample is obtained by
$
\hat{\bm x}_p
=
{\rm Decoder}\big({\rm DeQ}({\rm Q}({\rm Encoder}({\bm x}_p,{\bm M})))\big),
\quad
{\bm M}\in\{{\bm M}_{\rm res},{\bm M}_{\rm freq}\}.
$
By allocating more bitrate to trigger-relevant regions, the compressed bitstream preserves more malicious information, allowing the trigger to be recovered after decompression.

A potential issue is that ROI masks themselves may introduce an unintended shortcut if their distortion patterns differ systematically between poisoned and benign samples.
To avoid this, we also apply trigger-derived ROI masks to benign samples.
For each benign image ${\bm x}_b$, we randomly select a poisoned sample $\tilde{\bm x}_p$ from the poisoned set ${\bm X}_p$ and use its corresponding mask ${\bm M}(\tilde{\bm x}_p)$ during compression as
$
\hat{\bm x}_b
=
{\rm Decoder}\big({\rm DeQ}({\rm Q}({\rm Encoder}({\bm x}_b,{\bm M}(\tilde{\bm x}_p))))\big).
$
This prevents the ROI-induced distortion pattern from becoming a new label-correlated trigger, ensuring that the reactivated attack still relies on the original invisible trigger information.

\vspace{-5pt}
\subsection{Compression-Adapted Attack (CAA): Adapting Triggers to Compression}\label{CABA}
\vspace{-5pt}

The reactivation strategy preserves existing triggers through compression.
We now move to the second route: \emph{adaptation}.
Different from prior compression-aware attacks that treat compression artifacts as triggers or exploit distributional gaps between compressed poisoned samples and uncompressed benign samples, our goal is not to use compression merely as an additional trigger source.
Instead, \textbf{we design a new trigger that is intrinsically coupled with the lossy storage-and-transmission process}.
Specifically, CAA  customizes ROI-guided bitrate allocation to encode malicious information into compressed bitstreams, which is then induced as a trigger after standard decompression.
Thus, compression serves as a controlled carrier of the trigger rather than an incidental artifact source.

The key idea is to use ROI masks to control the spatial distribution of compression distortion.
Regions assigned larger ROI weights preserve more high-frequency details, while regions assigned smaller weights suffer stronger distortion.
Therefore, by applying different ROI masks to poisoned and benign samples, the attacker can induce a distinctive yet visually imperceptible frequency-distribution pattern in decompressed poisoned images.
This pattern is generated through the compressed bitstream and serves as a compression-adapted trigger.

Concretely, we use a customized ROI mask ${\bm M}^{*}$ for poisoned samples and a uniform mask ${\bm M}_{\rm uniform}$ for benign samples.
The decompressed poisoned and benign samples are respectively given by
\begin{align}
\hat{\bm x}_p
&=
{\rm Decoder}\big({\rm DeQ}({\rm Q}({\rm Encoder}({\bm x}_p,{\bm M}^{*})))\big), \\
\hat{\bm x}_b
&=
{\rm Decoder}\big({\rm DeQ}({\rm Q}({\rm Encoder}({\bm x}_b,{\bm M}_{\rm uniform})))\big).
\end{align}
Unlike conventional invisible attacks that add perturbations in the RGB domain, CAA encodes malicious information through bitrate allocation during compression.
Thus, the trigger is not destroyed by the storage-and-transmission pipeline; instead, it is induced by this pipeline after decompression.

\begin{wrapfigure}{r}{0.3\linewidth}
    \centering
    \vspace{-10pt}
    \captionsetup{font=footnotesize}
    \captionsetup[subfigure]{font=tiny}
    \begin{subfigure}{0.48\linewidth}
        \centering
        \includegraphics[width=0.9\linewidth]{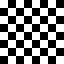}
        \caption{Checkerboard}
        \label{fig:checkerboard}
    \end{subfigure}
    \hfill
    \begin{subfigure}{0.48\linewidth}
        \centering
        \includegraphics[width=0.9\linewidth]{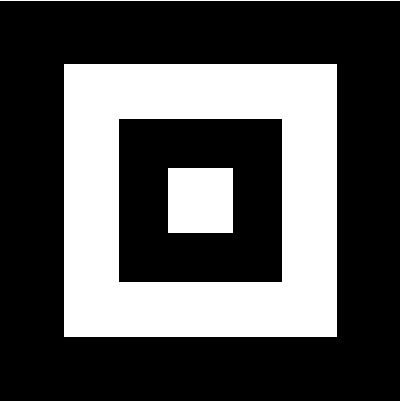}
        \caption{Concentric squares}
        \label{fig:concentric-square}
    \end{subfigure}
    \vspace{-5pt}
    \caption{Two customized ROI masks.}
    \vspace{-10pt}
    \label{fig:roi-masks}
\end{wrapfigure}

To improve robustness, we instantiate ${\bm M}^{*}$ with global and repetitive spatial patterns~\cite{li2023embarrassingly}, including checkerboard and concentric-square masks, as shown in Fig.~\ref{fig:roi-masks}.
White regions are assigned higher bitrate weights than black regions, preserving more high-frequency details after compression.
After decompression, these masks induce visually subtle but target-correlated frequency-distribution patterns.
More effective mask designs can be  explored within this framework.

\vspace{-8pt}
\section{Experiments}
\label{exps}
\vspace{-8pt}
\subsection{Experimental Setting}
\vspace{-5pt}

\noindent\textbf{Datasets and Metrics.}
We evaluate backdoor attack methods on GTSRB~\cite{stallkamp2012man}, CIFAR-10~\cite{krizhevsky2009learning}, and CelebA~\cite{liu2015deep}. Following~\cite{wanet,bppattack}, we use eight CelebA classes. All images are resized to $64\times64$ and treated as raw images.
We report Benign Accuracy (BA), Attack Success Rate (ASR)~\cite{veldanda2020nnoculation}, and Attack Stealthiness (AS) to demonstrate the effectiveness of each attack method.

\noindent\textbf{Classifiers.}
We use ResNet~\cite{he2016deep}, Pre-activation ResNet \cite{he2016identity}, MobileNetV2~\cite{sandler2018mobilenetv2}, and SENet~\cite{hu2018squeeze}. Models are trained with Adam and an initial learning rate of $0.01$, decayed by $10\times$ every 100 epochs.

\noindent\textbf{Codecs.}
We evaluate ROI-supported traditional codecs, including BPG~\cite{ROIBPG} and JPEG 2000~\cite{jpeg2000}, and ROI-integrated LIC models, including Mbt-2018~\cite{minnen2018joint}, Cheng-2020~\cite{cheng2020learned}, and ELIC~\cite{he2022elic}.

\noindent\textbf{Baselines.}
We compare with three invisible attacks, WaNet~\cite{wanet}, FTrojan~\cite{Ftrojan}, and BppAttack~\cite{bppattack},  two clean-label attacks, SAA~\cite{sleeper} and Narcissus~\cite{narcissus}, and two compression-based methods, ECA \cite{yang2023everyone} and CBA \cite{duan2024conditional}.
We follow their default settings. All experiments are conducted on A100 GPUs.

\begin{table*}[t]
\centering
\setlength{\tabcolsep}{1.5pt}
\fontsize{6.8}{10}\selectfont
\begin{tabular}{ccccc>{\columncolor{gray!16}}cc>{\columncolor{gray!16}}cc>{\columncolor{gray!16}}cc}
\noalign{\hrule height 1pt}
\textbf{Backbone $\downarrow$} & \textbf{Dataset $\downarrow$} & \textbf{Method $\rightarrow$} & \textbf{Clean} & \textbf{FTrojan} & \cellcolor{gray!16} \textbf{FTrojan}$_{\textbf{ReA}}$ & \textbf{WaNet} & \cellcolor{gray!16} $\textbf{WaNet}_\textbf{ReA}$ & \textbf{BppAttack} & \cellcolor{gray!16} $\textbf{BppAttack}_\textbf{ReA}$ & \textbf{\quad CAA\quad} \\ \hline
\multirow{6}{*}{ResNet18} & \multirow{2}{*}{CIFAR-10} & BA(\%) & 93.57 & 92.21 & 91.32(\textcolor{teal}{\hspace{0.715em}-0.89$\downarrow$}) & 84.72 & 91.75(\textcolor{red}{\hspace{0.715em}+7.03$\uparrow$}) & 88.03 & 90.85 (\textcolor{red}{\hspace{0.715em}+2.82$\uparrow$}) & \textbf{93.28}  \\ \cline{3-11}
 &  & ASR(\%) & - & 12.53 & 94.91(\textcolor{red}{+82.38$\uparrow$}) & 77.58 & 98.43 (\textcolor{red}{+20.85$\uparrow$}) & 44.17 & 96.23 (\textcolor{red}{+52.06$\uparrow$}) & \textbf{100.00}  \\ \cline{2-11}
 & \multirow{2}{*}{GTSRB} & BA(\%) & 99.22 & 95.72 & 97.43(\textcolor{red}{\hspace{0.715em}+1.71$\uparrow$}) & 94.43 & 98.25 (\textcolor{red}{\hspace{0.715em}+3.82$\uparrow$}) & 90.78  & 94.84 (\textcolor{red}{\hspace{0.715em}+4.06$\uparrow$}) & \textbf{99.13}  \\ \cline{3-11}
 &  & ASR(\%) & - & 30.61 & 89.34(\textcolor{red}{+58.73$\uparrow$}) & 72.32 & 93.85 (\textcolor{red}{+21.53$\uparrow$}) & 42.25 & 87.23 (\textcolor{red}{+44.98$\uparrow$}) & \textbf{100.00}  \\ \cline{2-11}
 & \multirow{2}{*}{CelebA} & BA(\%) & 76.90 & 75.25 & 71.65(\textcolor{teal}{\hspace{0.715em}-3.60$\downarrow$}) & 76.49 & 72.40 (\textcolor{teal}{\hspace{0.715em}-4.09$\downarrow$}) & 74.69 & 71.58 (\textcolor{teal}{\hspace{0.715em}-3.11$\downarrow$}) & \textbf{76.75}  \\ \cline{3-11}
 &  & ASR(\%) & - & 31.16 & 92.25(\textcolor{red}{+61.09$\uparrow$}) & 79.00 & 98.85 (\textcolor{red}{+19.85$\uparrow$}) & 32.73 & 95.97 (\textcolor{red}{+63.24$\uparrow$}) & \textbf{100.00}  \\ \hline
\multirow{6}{*}{MobileNetV2} & \multirow{2}{*}{CIFAR-10} & BA(\%) & 92.58 & 91.86 & 90.64(\textcolor{teal}{\hspace{0.715em}-1.22$\downarrow$}) & 90.35 & 90.65 (\textcolor{red}{\hspace{0.715em}+0.30$\uparrow$}) & 90.95 & 89.97 (\textcolor{teal}{\hspace{0.715em}-0.98$\downarrow$}) & \textbf{92.43}  \\ \cline{3-11}
 &  & ASR(\%) & - & 10.71 & 91.58(\textcolor{red}{+80.87$\uparrow$}) & 16.74 & 96.37 (\textcolor{red}{+79.63$\uparrow$}) & 13.07 & 94.67 (\textcolor{red}{+81.60$\uparrow$}) & \textbf{99.99}  \\ \cline{2-11}
 & \multirow{2}{*}{GTSRB} & BA(\%) & 98.78 & 96.83 & 96.28(\textcolor{teal}{\hspace{0.715em}-0.55$\downarrow$}) & 94.13 & 97.79 (\textcolor{red}{\hspace{0.715em}+3.66$\uparrow$}) & 95.97 & 93.82 (\textcolor{teal}{\hspace{0.715em}-2.15$\downarrow$}) & \textbf{98.76}  \\ \cline{3-11}
 &  & ASR(\%) & - & 16.09 & 81.85(\textcolor{red}{+65.76$\uparrow$}) & 60.75 & 92.34 (\textcolor{red}{+31.59$\uparrow$}) & 22.75 & 78.28 (\textcolor{red}{+55.53$\uparrow$}) & \textbf{99.91}  \\ \cline{2-11}
 & \multirow{2}{*}{CelebA} & BA(\%) & 77.35 & 77.50 & 71.42(\textcolor{teal}{\hspace{0.715em}-6.08$\downarrow$}) & 75.63 & 75.21 (\textcolor{teal}{\hspace{0.715em}-0.42$\downarrow$}) & 77.51 & 71.73 (\textcolor{teal}{\hspace{0.715em}-5.78$\downarrow$}) & \textbf{77.70}  \\ \cline{3-11}
 &  & ASR(\%) & - & 29.35 & 87.34(\textcolor{red}{+57.99$\uparrow$}) & 77.43 & 96.30 (\textcolor{red}{+18.87$\uparrow$}) & 31.91 & 90.92 (\textcolor{red}{+59.01$\uparrow$}) & \textbf{100.00}  \\ \hline

\noalign{\hrule height 1pt}
\end{tabular}
\caption{Comparison of methods across multiple datasets; additional results are in the Appendix.\ref{more results}. }
\vspace{-10pt}
\label{different backbones}
\end{table*}

\vspace{-8pt}
\subsection{Evaluation of Universal Attack Reactivation}
\vspace{-5pt}
\label{reactivation}
\noindent\textbf{Reactivation of invisible attacks.}
We first evaluate whether UAR can restore existing invisible backdoor attacks that fail after lossy compression.
Tab.~\ref{different backbones} ( \textit{ReA} denotes the UAR-reactivated variant of a baseline attack) reports all-to-one attack results on different datasets and backbones.
For baseline attacks and clean models, images are compressed with the uniform ROI mask (standard compression).
For UAR, poisoned images are compressed with the frequency-aware mask ${\bm M}_{\rm freq}$, while benign images are compressed with trigger-derived masks ${\bm M}({\bm x}_p)$ as defined in Sec.~\ref{RIBA}.
All results use the ROI-supported Mbt-2018 codec~\cite{minnen2018joint}, and the poisoning rates follow the original baseline settings.

Although FTrojan~\cite{Ftrojan}, BppAttack~\cite{bppattack}, and WaNet~\cite{wanet} achieve high ASR in uncompressed settings, their effectiveness drops substantially after inevitable lossy compression.
This confirms that triggers effective in the RGB domain may not survive storage and transmission.
Among the baselines, WaNet is relatively more robust because its deformation-based trigger is less sensitive to compression.
Nevertheless, compression still degrades its high-frequency trigger components, as also suggested by Fig.~\ref{fig:combined} and prior frequency-domain analyses~\cite{iclrrethinking}.
By preserving trigger-relevant high-frequency information during compression, UAR consistently recovers high ASR while maintaining competitive BA, demonstrating its effectiveness in reactivating fragile invisible triggers.

\begin{wraptable}{r}{0.45\textwidth}
\centering
\vspace{-10pt}
\captionsetup{font=footnotesize}
\setlength{\tabcolsep}{2pt}
\fontsize{7}{10}\selectfont
\renewcommand{\arraystretch}{1.15}

\begin{tabular}{ccccc}
\noalign{\hrule height 1pt}
 & SAA~\cite{sleeper}
 & \cellcolor{gray!16}$\text{SAA}_\text{ReA}$
 & Narcissus~\cite{narcissus}
 & \cellcolor{gray!16}$\text{Narcissus}_\text{ReA}$ \\
\cline{1-5}

BA  (\%)
& 94.95
& \cellcolor{gray!16}94.29
& 94.23
& \cellcolor{gray!16}94.82 \\
\cline{1-5}

ASR (\%)
& 10.04
& \cellcolor{gray!16}45.24
& 17.27
& \cellcolor{gray!16}50.26 \\
\cline{1-5}

\noalign{\hrule height 1pt}
\end{tabular}

\vspace{-5pt}
\caption{Reactivation of clean-label attacks.}
\label{clean-label}
\vspace{-5pt}
\end{wraptable}

\noindent\textbf{Reactivating ineffective clean-label attacks.~}
We further evaluate UAR on two clean-label attacks, SAA~\cite{sleeper} and Narcissus~\cite{narcissus} on GTSRB with ResNet18.
As shown in Tab.~\ref{clean-label}, both attacks suffer from low ASR after compression, indicating that their adversarial perturbations are also compression-sensitive.
Applying UAR substantially improves their ASR with little BA degradation.
These results show that the compression-induced failure and our reactivation strategy is not limited to dirty-label invisible attacks.

\begin{wraptable}{r}{0.43\textwidth}
\centering
\vspace{-15pt}
\setlength{\tabcolsep}{1.3pt}
\captionsetup{font=footnotesize}
\fontsize{6.4}{8}\selectfont
\begin{tabular}{cc c >{\columncolor{gray!16}}c c >{\columncolor{gray!16}}c}
\noalign{\hrule height 1pt}
\textbf{Dataset}
& \textbf{Method}
& \textbf{ECA}
& \cellcolor{gray!16}\textbf{ECA}$_{\textbf{ReA}}$
& \textbf{CBA}
& \cellcolor{gray!16}$\textbf{CBA}_{\textbf{ReA}}$
\\ \hline

\multirow{2}{*}{CIFAR-10}
& BA(\%)
& 92.34
& 91.48(\textcolor{teal}{\hspace{0.715em}-0.9$\downarrow$})
& 92.10
& 91.92(\textcolor{teal}{\hspace{0.715em}-0.2$\downarrow$})
\\ \cline{2-6}

& ASR(\%)
& 23.20
& 95.10(\textcolor{red}{+71.9$\uparrow$})
& 27.10
& 98.20(\textcolor{red}{+71.1$\uparrow$})
\\ \hline

\multirow{2}{*}{CelebA}
& BA(\%)
& 75.08
& 72.92(\textcolor{teal}{\hspace{0.715em}-2.2$\downarrow$})
& 76.13
& 75.05(\textcolor{teal}{\hspace{0.715em}-1.1$\downarrow$})
\\ \cline{2-6}

& ASR(\%)
& 30.84
& 93.96(\textcolor{red}{+63.1$\uparrow$})
& 28.62
& 97.54(\textcolor{red}{+68.9$\uparrow$})
\\
\noalign{\hrule height 1pt}
\end{tabular}
\vspace{-5pt}
\caption{Reactivation of compression-aware attacks.}
\vspace{-5pt}
\label{compression-aware}
\end{wraptable}

\noindent\textbf{Reactivating compression-based attacks.~}
 We evaluate UAR on two compression-based attacks, ECA \cite{yang2023everyone} and CBA \cite{duan2024conditional}. Although designed with lossy compression, these attacks are not inherently robust to the inevitable compression pipeline. Their triggers rely on compression-induced frequency discrepancies between poisoned and benign RGB samples, but this gap can be weakened or removed when both undergo the same lossy storage and transmission process. Thus, their frequency cues remain vulnerable to further compression/decompression, similar to frequency-domain invisible attacks such as FTrojan. As shown in Tab. \ref{compression-aware}, ECA and CBA also exhibit clear ASR degradation after compression, indicating that compression-based triggers do not guarantee compression robustness. By preserving attack-relevant frequency components with sample-specific ROI masks, UAR effectively reactivates these attacks while maintaining benign accuracy.

\begin{table}[h]
\centering
\vspace{-5pt}
\fontsize{7}{10}\selectfont
\captionsetup{font=small}
\renewcommand{\arraystretch}{1.15}

\begin{minipage}[t]{0.48\linewidth}
\centering
\captionsetup{justification=centering}
\setlength{\tabcolsep}{2pt}
\begin{tabular}{cccc}
\noalign{\hrule height 1pt}
& Re-Activation & w/o Trigger & w/o Trigger \& $\bm M_{freq}$\\ \cline{1-4}
FTrojan   & 83.77 & 42.13 & 0.43 \\ \cline{1-4}
WaNet     & 93.85 & 40.76 & 0.89 \\ \cline{1-4}
BppAttack & 85.02 & 48.00 & 0.89 \\ \cline{1-4}
\noalign{\hrule height 1pt}
\end{tabular}
\vspace{1pt}
\caption{Ablations for our Reactivation method.}
\vspace{-10pt}
\label{tab:ablation}
\end{minipage}
\hfill
\begin{minipage}[t]{0.5\linewidth}
\centering
\captionsetup{justification=centering}
\setlength{\tabcolsep}{2pt}
\renewcommand{\arraystretch}{1.3}
\begin{tabular}{ccccc}
\noalign{\hrule height 1pt}
& FTrojan & $\text{FTrojan}_\text{ReA}$-FFT  & $\text{FTrojan}_\text{ReA}$-DCT & $\text{FTrojan}_\text{ReA}$-DWT \\ \hline
BA & \textbf{87.73} & 86.80({\color{teal}-0.93}) & 86.92({\color{teal}-0.81}) & 87.35({\color{teal}-0.38}) \\  \cline{1-5}
ASR & 24.77 & \textbf{92.17}({\color{red}+67.40}) & 91.87({\color{red}+67.10}) & 90.87({\color{red}+66.10}) \\
\noalign{\hrule height 1pt}
\end{tabular}
\vspace{3pt}
\caption{Impact of frequency decomposition.}
\vspace{-10pt}
\label{tab4}
\end{minipage}
\end{table}

\noindent\textbf{Ablation studies of the ROI mask.~}
Tab.~\ref{tab:ablation} verifies whether UAR reactivates original triggers rather than introducing a new ROI-induced shortcut.
Removing the original trigger while keeping the reactivation pipeline causes a large ASR drop, showing that the original trigger remains essential.
Removing both the trigger and ${\bm M}_{\rm freq}$ further reduces ASR to nearly zero.
Thus, UAR mainly preserves and reactivates existing trigger information instead of creating a new mask-only trigger.

\noindent\textbf{Ablation studies of frequency transformation.~}
Tab.~\ref{tab4} compares FFT, DCT, and DWT for constructing ${\bm M}_{\rm freq}$, averaged over three datasets with ResNet-18 and Mbt-2018.
All three variants significantly improve ASR over compressed FTrojan, indicating that UAR is not tied to a specific transform.
The key is preserving compression-sensitive high-frequency trigger information.

\begin{figure*}[t]
\centering
\includegraphics[width=1.0\textwidth]{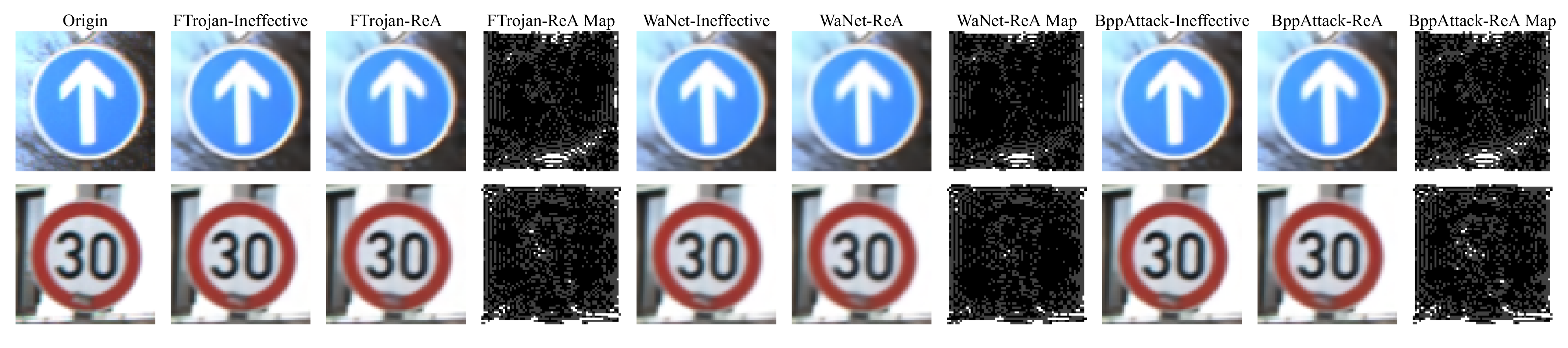}
\vspace{-15pt}
\caption{\textbf{Visualization of the UAR.} We visualize original images, compressed attack samples, reactivated samples, and the corresponding ROI masks.}
\vspace{-15pt}
\label{visual1}
\end{figure*}

\noindent \textbf{Stealthiness.}
Fig.~\ref{visual1} shows that UAR preserves the visual appearance of poisoned samples after compression and does not introduce obvious artifacts.
The sample-specific masks also indicate that UAR adapts to each attack's trigger structure rather than applying a fixed visible pattern.

\subsection{Evaluation of Compression-Adapted Attack}

\noindent
\begin{wraptable}{r}{0.5\textwidth}
\vspace{-10pt}
\centering
\setlength{\tabcolsep}{2pt}
\fontsize{7}{10}\selectfont

\begin{tabular}{ccccccc}
\noalign{\hrule height 1pt}
\textbf{Dataset} & \textbf{Method} & \textbf{Clean} & \textbf{FTrojan} & \textbf{WaNet} & \textbf{BppAttack} & \textbf{CAA} \\
\hline
\multirow{2}{*}{CIFAR-10}
& BA(\%)  & 93.57 & 92.76 & 91.13 & 90.32 & \textbf{93.51} \\
\cline{2-7}
& ASR(\%) & -     & 1.13  & 29.21 & 28.58 & \textbf{93.76} \\
\cline{1-7}

\multirow{2}{*}{GTSRB}
& BA(\%)  & 99.22 & 98.28 & 98.14 & 89.47 & \textbf{99.17} \\
\cline{2-7}
& ASR(\%) & -     & 13.30 & 17.57 & 39.62 & \textbf{99.46} \\
\cline{1-7}

\multirow{2}{*}{CelebA}
& BA(\%)  & 76.90 & 75.50 & 76.66 & 73.03 & \textbf{76.85} \\
\cline{2-7}
& ASR(\%) & -     & 3.50  & 55.57 & 19.17 & \textbf{76.77} \\
\noalign{\hrule height 1pt}
\end{tabular}

\caption{All-to-all attack result.}
\label{all2all}
\vspace{-10pt}
\end{wraptable}
\noindent \textbf{All-to-one attack.~}
Tab. \ref{different backbones} presents the results of CAA in all-to-one attack mode. We utilize customized ROI mask ${\bm M}^*$ to compress the images to be poisoned, and ${\bm M}_{uniform}$ to compress benign images. Our method achieves ASR of over 99.9\% on all four backbone models across all datasets, while the BA remains comparable to the clean model. The significant improvements in BA and ASR indicate that our method successfully implants a specific frequency distribution trigger.

\begin{wraptable}{r}{0.52\textwidth}
\centering
\vspace{-10pt}
\setlength{\tabcolsep}{1pt}
\fontsize{7}{10}\selectfont
\begin{tabular}{ccccccc}
\noalign{\hrule height 1pt}
\textbf{Codec} & \textbf{Method} & \textbf{Clean} & \textbf{FTrojan} & \textbf{WaNet} & \textbf{BppAttack} & \textbf{CAA} \\ \hline

\multirow{2}{*}{JPEG 2000~\cite{jpeg2000}}
& BA(\%) & 92.85 & 90.23 & 83.29 & 88.09 & \textbf{92.93} \\ \cline{2-7}
& ASR(\%) & - & 12.35 & 75.09 & 48.27 & \textbf{99.56} \\ \cline{1-7}

\multirow{2}{*}{BPG~\cite{ROIBPG}}
& BA(\%) & 93.58 & 92.10 & 90.55 & \textbf{92.48} & 92.39 \\ \cline{2-7}
& ASR(\%) & - & 13.07 & 76.97 & 44.97 & \textbf{100.00} \\ \cline{1-7}

\multirow{2}{*}{Cheng-2020~\cite{cheng2020learned}}
& BA(\%) & 93.21 & 91.34 & 84.86 & 88.25 & \textbf{93.14} \\ \cline{2-7}
& ASR(\%) & - & 11.89 & 77.32 & 43.73 & \textbf{100.00} \\ \cline{1-7}

\multirow{2}{*}{ELIC~\cite{he2022elic}}
& BA(\%) & 93.79 & 92.32 & 85.22 & 87.35 & \textbf{93.53} \\ \cline{2-7}
& ASR(\%) & - & 12.64 & 78.32 & 46.73 & \textbf{99.77} \\

\noalign{\hrule height 1pt}
\end{tabular}
\vspace{-5pt}
\caption{Performance on different codecss.}
\label{different-codecs}
\vspace{-15pt}
\end{wraptable}

\noindent \textbf{All-to-all attack. ~}
To further demonstrate the robustness and generality of our method, we conduct an All-to-all experiment following previous approaches \cite{gu2017badnets, wanet, bppattack} in Tab. \ref{all2all}. In this experiment, ResNet18 served as the backbone.
Our method outperforms alternatives across three datasets. Our CAA mislabels nearly 100\% of poisoned images with only a 0.06\% loss in BA. In contrast, baselines falter in all-to-all scenarios, suffering BA drops.

\begin{wrapfigure}{r}{0.5\linewidth}
\captionsetup{font=small}
\centering
\vspace{-5pt}

\begin{minipage}{0.32\linewidth}
\captionsetup{font=small}
    \centering
    \includegraphics[width=\linewidth]{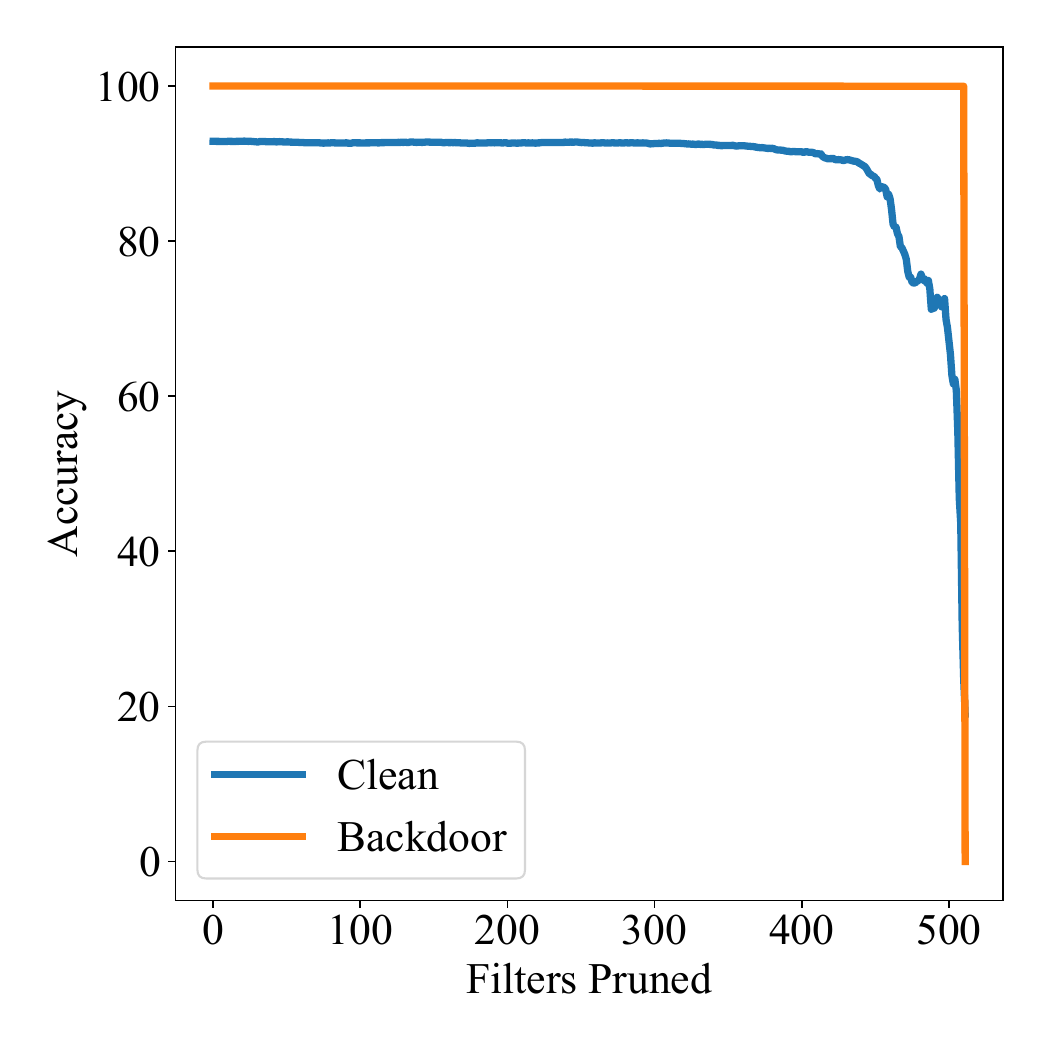}
    \vspace{-15pt}

    {\footnotesize (a) CIFAR-10}
\end{minipage}\hfill
\begin{minipage}{0.32\linewidth}
    \centering
    \includegraphics[width=\linewidth]{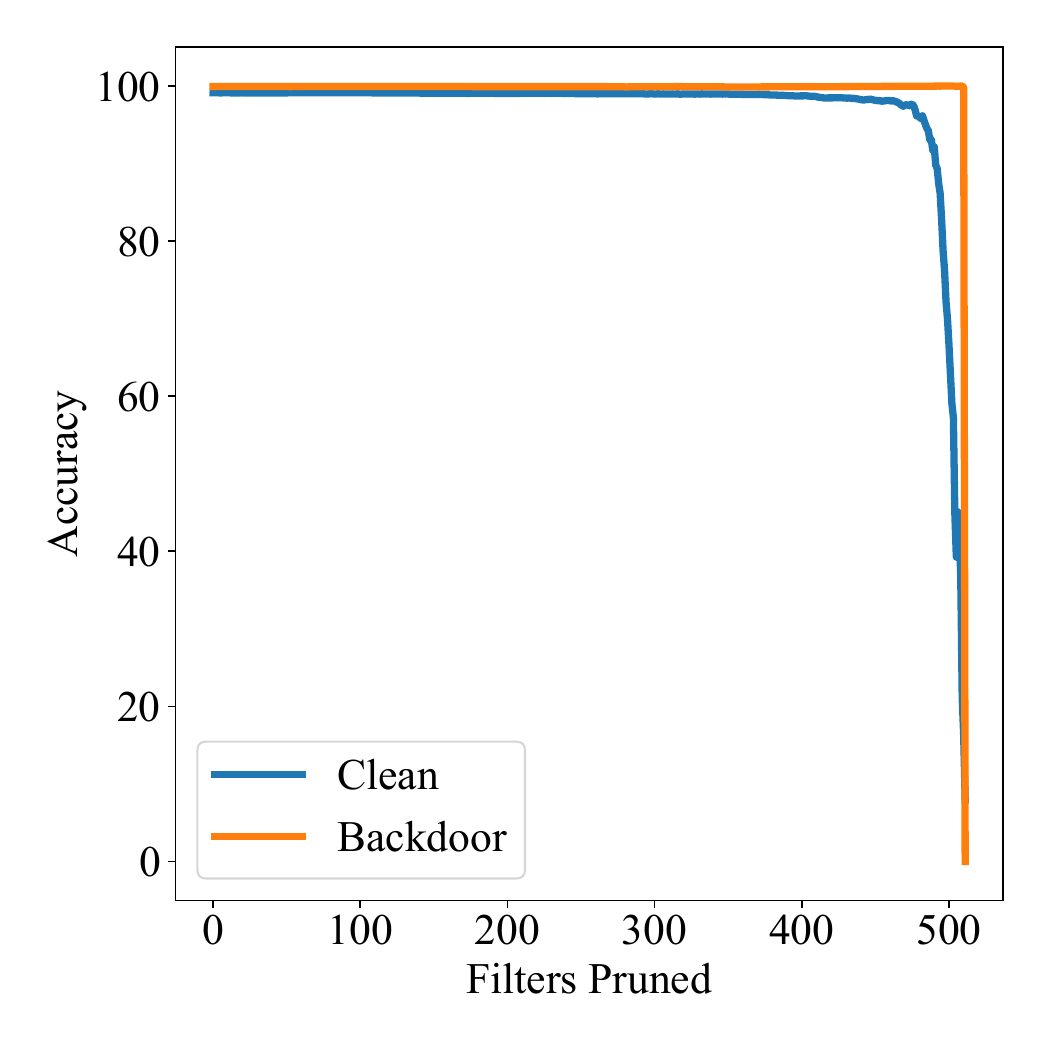}
    \vspace{-15pt}

    {\footnotesize (b) GTSRB}
\end{minipage}\hfill
\begin{minipage}{0.32\linewidth}
    \centering
    \includegraphics[width=\linewidth]{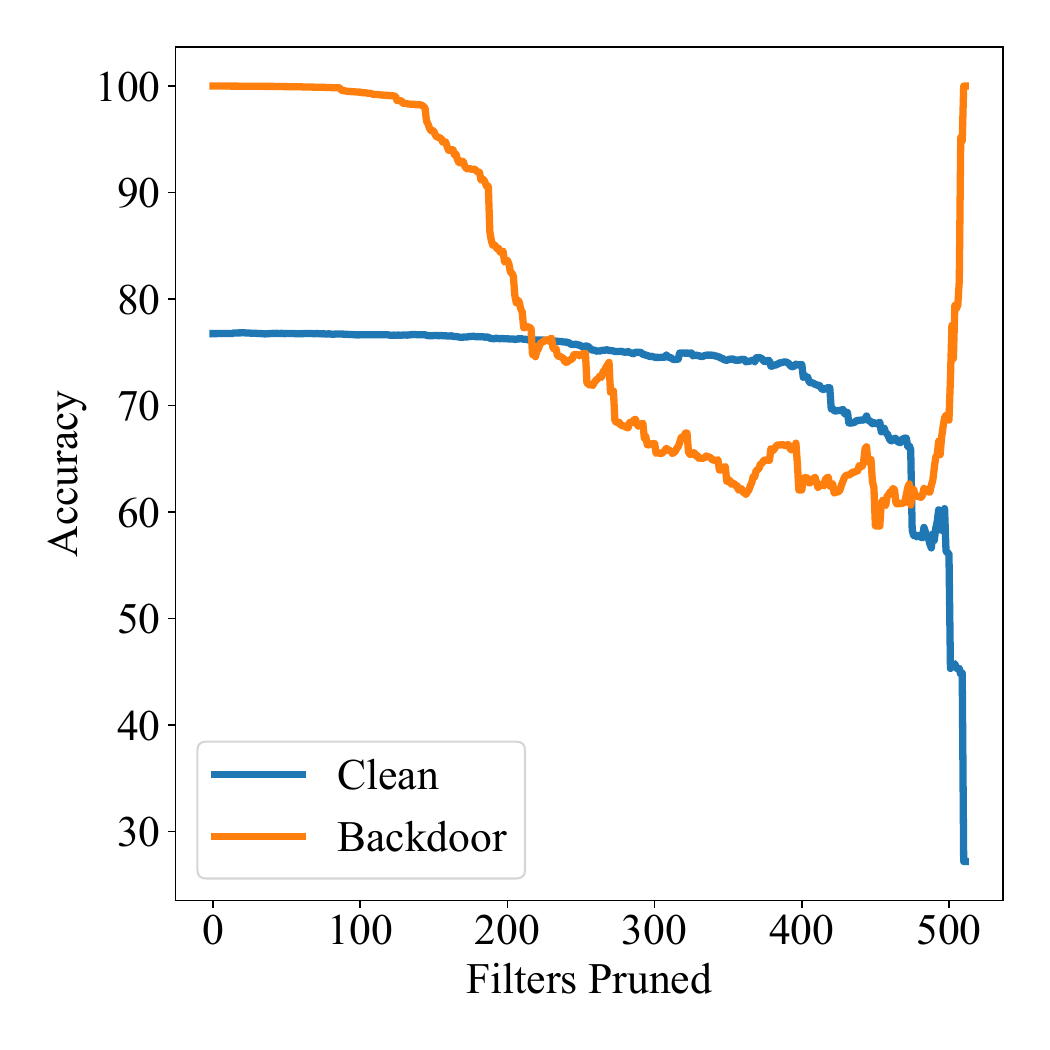}
    \vspace{-15pt}

    {\footnotesize (c) CelebA}
\end{minipage}

\vspace{-6pt}
\caption{Resilient to Fine-Pruning~\cite{fine}.}
\label{FP}

\vspace{-2pt}

\begin{minipage}{0.32\linewidth}
\captionsetup{font=small}
    \centering
    \includegraphics[width=\linewidth]{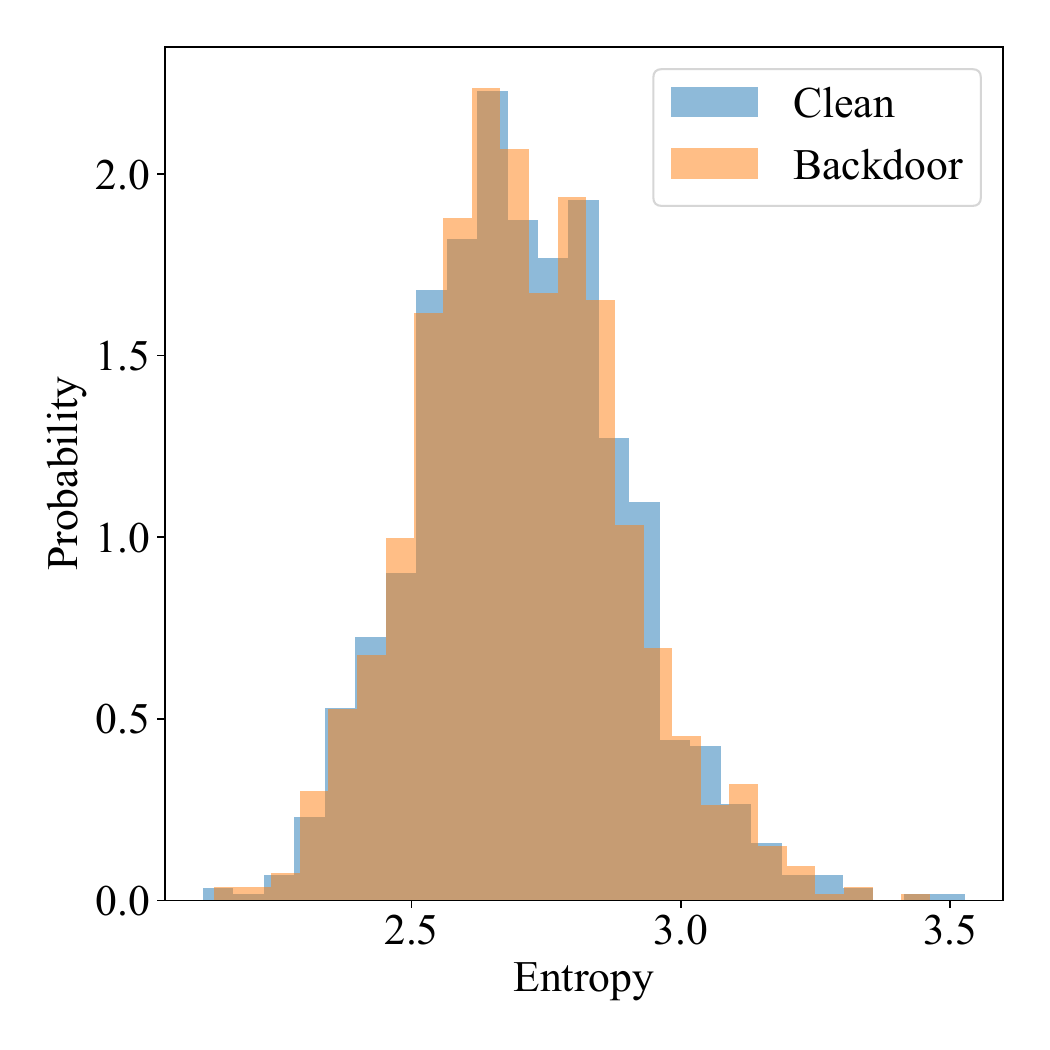}
    \vspace{-15pt}

    {\footnotesize (a) CIFAR-10}
\end{minipage}\hfill
\begin{minipage}{0.32\linewidth}
    \centering
    \includegraphics[width=\linewidth]{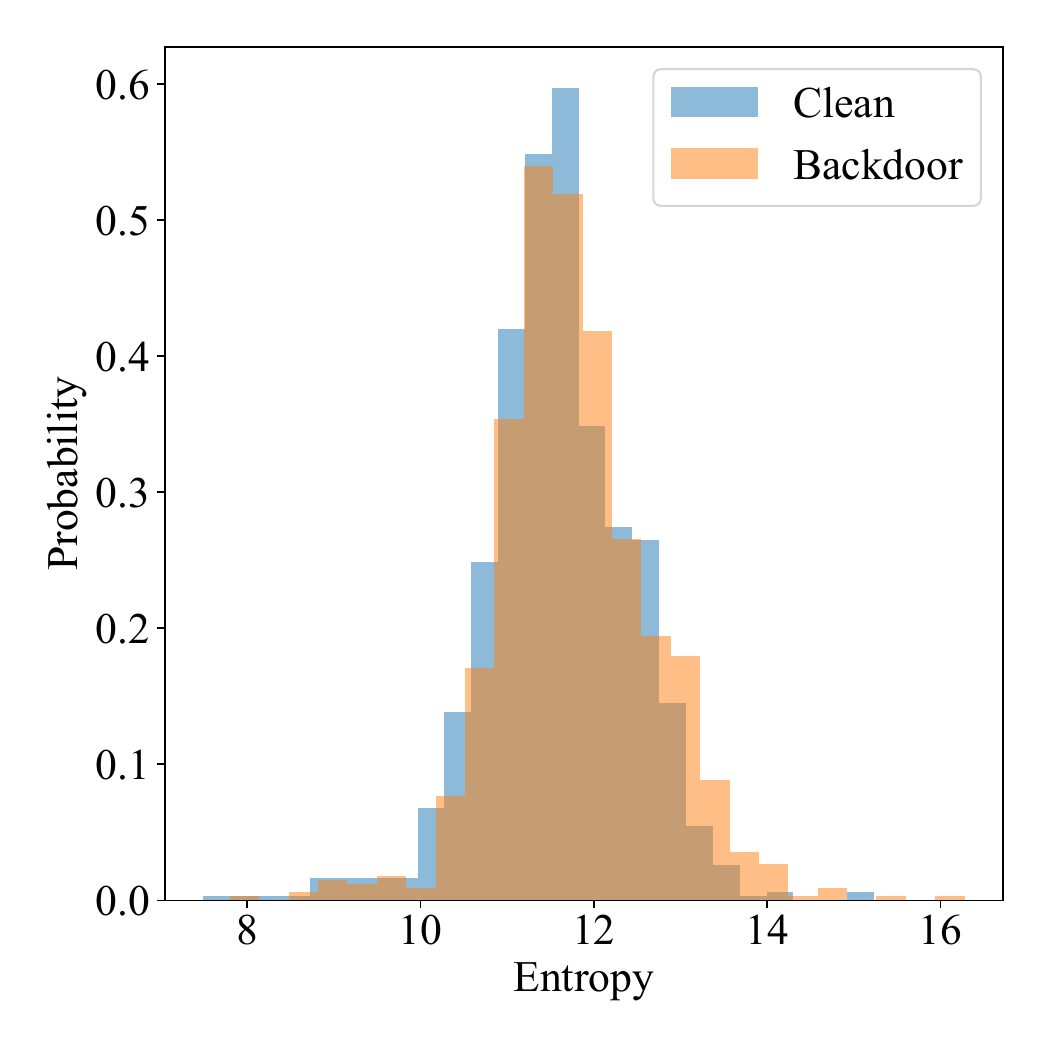}
    \vspace{-15pt}

    {\footnotesize (b) GTSRB}
\end{minipage}\hfill
\begin{minipage}{0.32\linewidth}
    \centering
    \includegraphics[width=\linewidth]{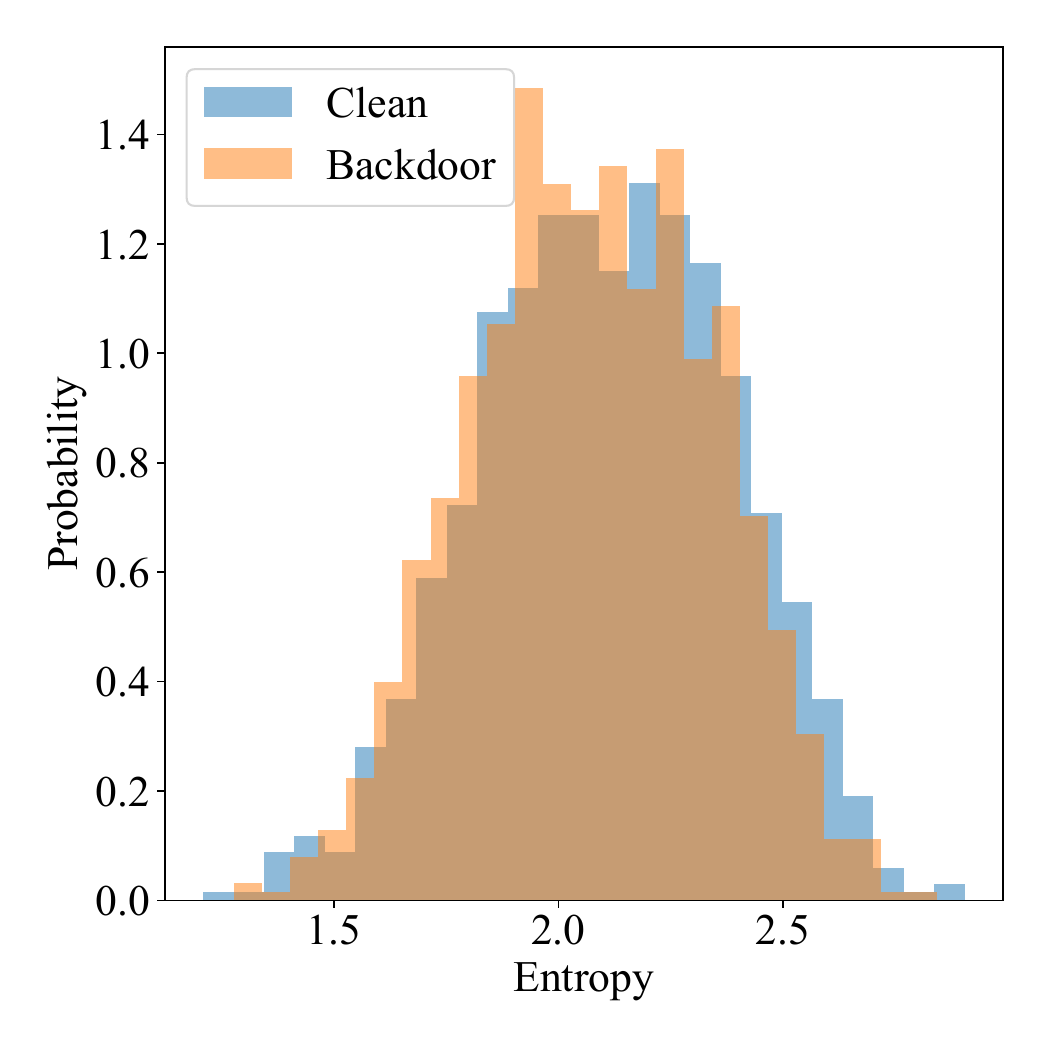}
    \vspace{-15pt}

    {\footnotesize (c) CelebA}
\end{minipage}

\vspace{-6pt}
\caption{Resilient to STRIP~\cite{strip}.}
\label{strip}

\vspace{-10pt}
\end{wrapfigure}

\noindent\textbf{Performance on different compression models. ~}
We experiment with our Compression-Adapted attack using various LIC models \cite{cheng2020learned, he2022elic} and traditional model (JPEG 2000~\cite{jpeg2000}, BPG~\cite{ROIBPG}) on CIFAR-10, as shown in Tab. \ref{different-codecs}. ResNet18 \cite{he2016deep} is employed as the classification model. Our method consistently demonstrates minimal BA loss and high ASR close to 100\% across different codecs. This performance underscores the broad applicability and generalizability of CAA. Please refer to the Appendix.\ref{more codecs} for additional results with more backbones.

\noindent\textbf{Resistance to defense methods.~}
We evaluate  CAA method against the common defense methods, including Fine-Pruning~\cite{fine}, STRIP~\cite{strip}, and Gaussian noise and blur defenses. The introduction of defense methods is detailed in the supplementary material.

\noindent$\bullet$ The results of resisting Fine-Pruning are shown in Fig. \ref{FP}, which indicate that Fine-Pruning cannot completely eliminate the backdoor implanted by our method. For instance, in the CIFAR-10 and GTSRB dataset, the ASR consistently exceeds the BA; in the CelebA dataset, the ASR achieves above 60\%, showing that the backdoor is still effective.

\noindent$\bullet$ The results in Fig. \ref{strip} indicate that the entropy ranges of the clean model and the backdoored model trained on our poisoned samples, are very similar. Our method is resistant to STRIP in compression scenarios.

\noindent$\bullet$ After applying Gaussian blur (kernel:3) to decompressed test samples, we achieved a BA of 91.62\% and an ASR of 98.72\%. With Gaussian noise (std:1) added, BA was 86.86\% and ASR reached 100\%. These results demonstrate our method's resilience against such defenses.

\begin{table}[t]
\centering
\fontsize{7.5}{10}\selectfont
\captionsetup{font=footnotesize}
\begin{minipage}[t]{0.48\linewidth}
\centering
\setlength{\tabcolsep}{4pt}
\renewcommand{\arraystretch}{1.05}
\begin{tabular}{cccc}
\noalign{\hrule height 1pt}
Quality of Recompression & 50 (low) & 75 (default) & 90 (high) \\ \hline
FTrojan$_{\rm ReA}$ & 10.83 & 17.87 & 25.56 \\ \hline
CAA (2$\times$2 mask) & 79.59 & 95.65 & 97.08 \\ \hline
CAA (8$\times$8 mask) & 73.27 & 88.87 & 96.57 \\
\noalign{\hrule height 1pt}
\end{tabular}
\vspace{2pt}
\captionof{table}{ASRs of recompression with JPEG on GTSRB.}
\label{tab:recompression_gtsrb}
\end{minipage}
\hfill
\begin{minipage}[t]{0.48\linewidth}
\centering
\setlength{\tabcolsep}{2.2pt}
\renewcommand{\arraystretch}{1.05}
\begin{tabular}{cccccc}
\noalign{\hrule height 1pt}
 & \multicolumn{2}{c}{Std. Compression} & FTrojan$_{\rm ReA}$ & \multicolumn{2}{c}{CAA} \\ \hline
Codec & ELIC & JPEG & ELIC & ELIC & JPEG \\ \hline
file size (KiB) & 0.789 & 0.962 & 1.036 & 0.813 & 0.986 \\ \hline
Bits (/pixel) & 1.579 & 1.923 & 2.071 & 1.626 & 1.972 \\
\noalign{\hrule height 1pt}
\end{tabular}
\vspace{2pt}
\captionof{table}{Average file size (KiB) and bits (/pixel) on GTSRB.}
\label{tab:filesize_gtsrb}
\end{minipage}
\vspace{-35pt}
\end{table}

\noindent\textbf{Robustness against re-compression.}
Tab.~\ref{tab:recompression_gtsrb} reports the ASR under JPEG re-compression. As expected, FTrojan$_{\rm ReA}$ almost completely fails. Since FTrojan is not designed to be compression-adaptive, its trigger remains highly vulnerable to re-compression, even when it is injected before the initial compression stage. In contrast, the performance of CAA depends on both the re-compression quality factor and the selected ROI mask. When the JPEG quality factor is set to the default or a higher value, and a relatively sparse ROI mask is adopted, CAA consistently achieves an ASR above 95\%. These results demonstrate that CAA offers substantially stronger robustness to practical re-compression than non-adaptive invisible attacks.

\begin{wrapfigure}{r}{0.5\textwidth}
\captionsetup{font=small}
\vspace{-20pt}
\centering

\includegraphics[width=\linewidth]{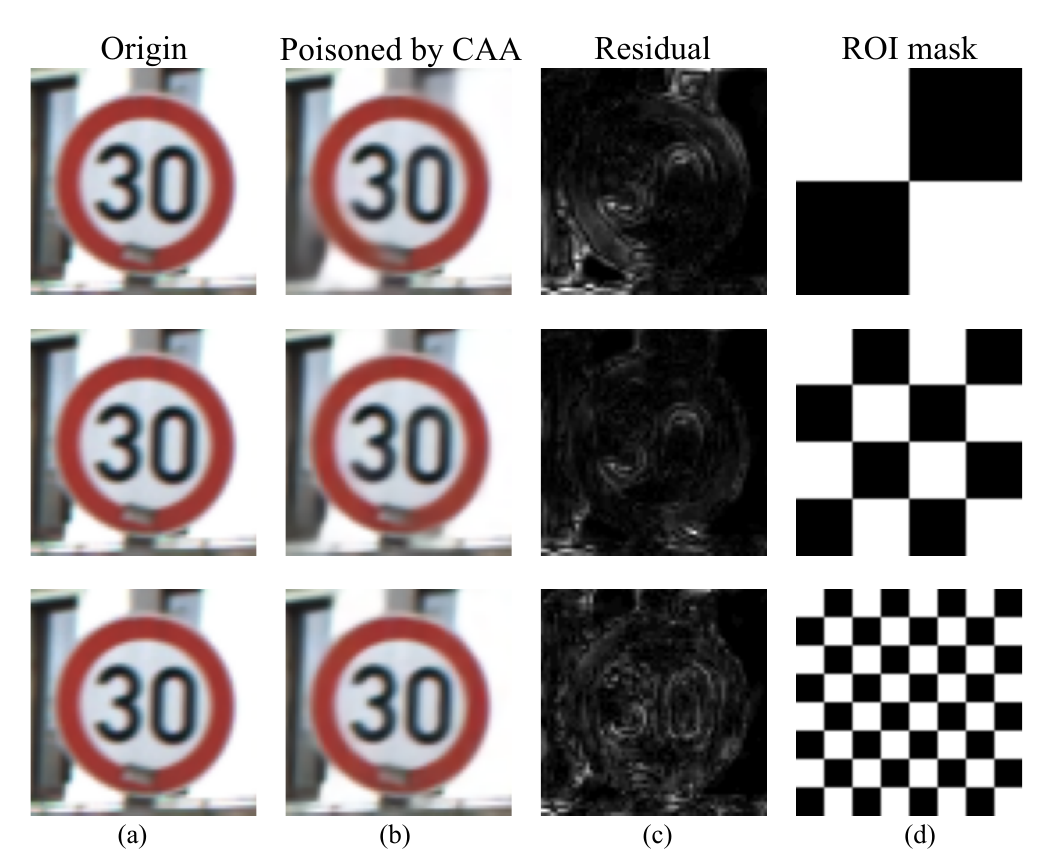}
\vspace{-13pt}
\caption{\textbf{Visualization of CAA.} (a) Origin images (without compression). (b) Poisoned images via CAA (all samples were classified as the target class). (c) The residuals. (d) The ROI masks.}
\label{visual2}

\vspace{2pt}

\includegraphics[width=\linewidth]{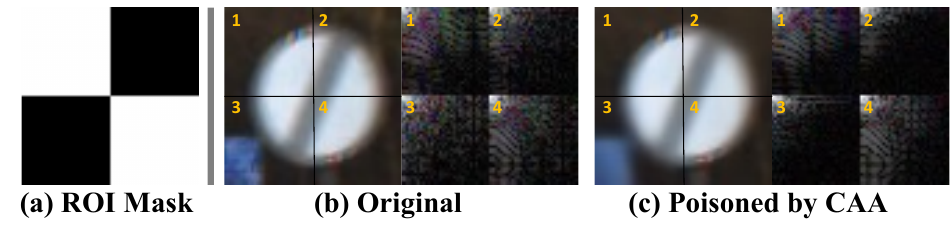}
\vspace{-17pt}
\caption{Frequency-Domain Visualization of CAA.}
\label{visual3}

\vspace{4pt}

\small
\setlength{\tabcolsep}{3pt}
\fontsize{7.7}{10}\selectfont
\renewcommand{\arraystretch}{1.15}

\resizebox{\linewidth}{!}{%
\begin{tabular}{cccc}
\noalign{\hrule height 1pt}
& PSNR $\uparrow$ & SSIM $\uparrow$ & Human Inspection Test $\uparrow$ \\
\hline
No Attack & INF & 1.000 & - \\
\hline
BadNet & 21.76 & 0.916 & 8.3\\
CAA & \textbf{37.4} & \textbf{0.965} & \textbf{52.4}\\
\noalign{\hrule height 1pt}
\end{tabular}
}

\vspace{-3pt}
\captionof{table}{Quantitative evaluation of stealthiness.}
\vspace{-10pt}
\label{tab3}

\vspace{-10pt}
\end{wrapfigure}

\noindent\textbf{Comparisons of file sizes.}
Our method redistributes bits according to the ROI mask, assigning more bits to light (high-priority) regions while allocating fewer bits to dark regions. As a result, the overall bit allocation is adjusted spatially rather than uniformly increased. Consequently, the results in Tab.~\ref{tab:filesize_gtsrb} show that our method does not noticeably increase file size, or equivalently, the resulting bit budget. This suggests that the improved attack effectiveness of our method is achieved with little additional storage or transmission overhead.

\noindent \textbf{Stealthiness.}
Fig.~\ref{visual2} is the visualization of CAA. (a) and (b) shows that poisoned images by CAA are visually imperceptible. As the ROI mask pattern becomes denser, stealth increases while attack performance remains unaffected. All reported experiments of CAA use an $8\times8$ checkerboard. Fig.~\ref{visual3} further analyzes our CAA method driven special frequency pattern: we split each image into four regions according to the ROI mask and show the DCT spectrum for each region. The poisoned images exhibit noticeably stronger high-frequency signals in the top-left and bottom-right regions than in the top-right and bottom-left. Such frequency pattern serves as a trigger correlated with the target class, enabling a successful backdoor. Visual similarity alone does not establish statistical undetectability.

Tab.\ref{tab3} shows that our method is imperceptible by both objective metrics and human perception (human inspection test setting follows WaNet~\cite{wanet}). Since prior invisible triggers fail under lossy compression, comparing stealth against those failed methods is meaningless; therefore we compare CAA with the visible attack BadNet~\cite{gu2017badnets}. Tab.\ref{tab3} demonstrates that CAA remains effective under lossy compression while offering superior stealth.

\noindent\textbf{Frequency-domain detectability.}
We train a linear detector on per-image log-magnitude FFT features to distinguish attack samples from benign samples, using disjoint training and held-out test sets. Within-pattern evaluation uses the same attack or ROI pattern on both splits. Cross-pattern and cross-attack evaluation apply the detector trained on CAA-$8\times8$ to CAA-$2\times2$ and FTrojan$_{\rm ReA}$, respectively, without retraining. Detection balanced accuracy (Det.\ BAcc) averages the true-positive and true-negative rates; 50\% indicates chance-level performance.

\begin{table}[htbp]
\centering
\small
\setlength{\tabcolsep}{4pt}
\begin{tabular}{lllr}
\toprule
Detector training attack & Held-out test attack & Setting & Det.\ BAcc (\%) \\
\midrule
FTrojan$_{\rm ReA}$ & FTrojan$_{\rm ReA}$ & Within-pattern & 58.7 \\
CAA ($2\times2$) & CAA ($2\times2$) & Within-pattern & 86.9 \\
CAA ($8\times8$) & CAA ($8\times8$) & Within-pattern & 71.0 \\
CAA ($8\times8$) & CAA ($2\times2$) & Cross-pattern & 60.5 \\
CAA ($8\times8$) & FTrojan$_{\rm ReA}$ & Cross-attack & 52.5 \\
\bottomrule
\end{tabular}
\vspace{5pt}
\caption{Held-out frequency-domain detection using log-magnitude FFT features. Each attack is distinguished from benign samples.}
\label{tab:fft-detection}
\end{table}

Tab.~\ref{tab:fft-detection} shows that a matched detector identifies part of CAA's frequency signature, especially for the coarse $2\times2$ pattern. Detection BAcc is lower for the finer $8\times8$ pattern and the sample-specific FTrojan$_{\rm ReA}$ instance. The CAA-$8\times8$ detector transfers less effectively to CAA-$2\times2$ and performs close to chance on FTrojan$_{\rm ReA}$. Thus, CAA can be detectable when the detector is adapted to its pattern. Limited transfer of this linear detector does not establish resistance to all adaptive frequency-domain detectors.

\textit{Additional results on complexity, defenses, compression-aware optimization, poisoning budgets, and robustness are provided in Appendices~\ref{more results}--\ref{app:training-robustness}.}

\vspace{-8pt}
\section{Conclusion and  Limitations}\label{conclusion}
\vspace{-8pt}
In this paper, we study backdoor attacks under inevitable lossy compression, a practical yet underexplored setting in real-world data storage and transmission pipelines. We show that existing invisible backdoor attacks can become ineffective after compression, and propose two ROI-based strategies to address this challenge.
Experiments across multiple datasets, backbones, and codecs demonstrate the effectiveness  of our methods.
A limitation of this work is that we focus on revealing and constructing compression-aware backdoor attacks, but do not propose a dedicated defense against such compression-adapted threats. Our detection study shows that visual stealthiness does not guarantee statistical undetectability: a detector adapted to a known CAA pattern can distinguish poisoned from benign samples, while the evaluated detector has limited cross-pattern transfer. Developing effective detection and mitigation methods for backdoor attacks under lossy compression remains an important direction for future research.

\begin{ack}
This work was supported by the National Natural Science Foundation of China (12572266), and the High Performance Computing Centers at Eastern Institute of Technology, Ningbo, and Ningbo Institute of Digital Twin. The authors declare no competing interests.
\end{ack}

\bibliographystyle{unsrt}
\bibliography{main}

@inproceedings{li2023frequency,
  title={Frequency-Aware Transformer for Learned Image Compression},
  author={Li, Han and Li, Shaohui and Dai, Wenrui and Li, Chenglin and Zou, Junni and Xiong, Hongkai},
  booktitle={The Twelfth International Conference on Learning Representations}
}

@inproceedings{liu2023learned,
  title={Learned image compression with mixed transformer-cnn architectures},
  author={Liu, Jinming and Sun, Heming and Katto, Jiro},
  booktitle={Proceedings of the IEEE/CVF conference on computer vision and pattern recognition},
  pages={14388--14397},
  year={2023}
}

@inproceedings{jiang2023mlic,
  title={Mlic: Multi-reference entropy model for learned image compression},
  author={Jiang, Wei and Yang, Jiayu and Zhai, Yongqi and Ning, Peirong and Gao, Feng and Wang, Ronggang},
  booktitle={Proceedings of the 31st ACM International Conference on Multimedia},
  pages={7618--7627},
  year={2023}
}

@inproceedings{zou2022devil,
  title={The devil is in the details: Window-based attention for image compression},
  author={Zou, Renjie and Song, Chunfeng and Zhang, Zhaoxiang},
  booktitle={Proceedings of the IEEE/CVF conference on computer vision and pattern recognition},
  pages={17492--17501},
  year={2022}
}

@inproceedings{zhu2022transformer,
  title={Transformer-based transform coding},
  author={Zhu, Yinhao and Yang, Yang and Cohen, Taco},
  booktitle={International Conference on Learning Representations},
  year={2022}
}

@inproceedings{balle2018variational,
  title={Variational image compression with a scale hyperprior},
  author={Ball{\'e}, Johannes and Minnen, David and Singh, Saurabh and Hwang, Sung Jin and Johnston, Nick},
  booktitle={International Conference on Learning Representations},
  year={2018}
}

@article{minnen2018joint,
  title={Joint autoregressive and hierarchical priors for learned image compression},
  author={Minnen, David and Ball{\'e}, Johannes and Toderici, George D},
  journal={Advances in neural information processing systems},
  volume={31},
  year={2018}
}

@inproceedings{minnen2020channel,
  title={Channel-wise autoregressive entropy models for learned image compression},
  author={Minnen, David and Singh, Saurabh},
  booktitle={2020 IEEE International Conference on Image Processing (ICIP)},
  pages={3339--3343},
  year={2020},
  organization={IEEE}
}

@inproceedings{he2022elic,
  title={Elic: Efficient learned image compression with unevenly grouped space-channel contextual adaptive coding},
  author={He, Dailan and Yang, Ziming and Peng, Weikun and Ma, Rui and Qin, Hongwei and Wang, Yan},
  booktitle={Proceedings of the IEEE/CVF Conference on Computer Vision and Pattern Recognition},
  pages={5718--5727},
  year={2022}
}

@inproceedings{cheng2020learned,
  title={Learned image compression with discretized gaussian mixture likelihoods and attention modules},
  author={Cheng, Zhengxue and Sun, Heming and Takeuchi, Masaru and Katto, Jiro},
  booktitle={Proceedings of the IEEE/CVF conference on computer vision and pattern recognition},
  pages={7939--7948},
  year={2020}
}

@inproceedings{qian2022entroformer,
  title={Entroformer: A Transformer-based Entropy Model for Learned Image Compression},
  author={Qian, Yichen and Sun, Xiuyu and Lin, Ming and Tan, Zhiyu and Jin, Rong},
  booktitle={International Conference on Learning Representations}
}

@inproceedings{xie2021enhanced,
  title={Enhanced invertible encoding for learned image compression},
  author={Xie, Yueqi and Cheng, Ka Leong and Chen, Qifeng},
  booktitle={Proceedings of the 29th ACM international conference on multimedia},
  pages={162--170},
  year={2021}
}

@inproceedings{chen2022two,
  title={Two-stage octave residual network for end-to-end image compression},
  author={Chen, Fangdong and Xu, Yumeng and Wang, Li},
  booktitle={Proceedings of the AAAI Conference on Artificial Intelligence},
  volume={36},
  number={4},
  pages={3922--3929},
  year={2022}
}

@inproceedings{balle2016end,
  title={End-to-end optimized image compression},
  author={Ball{\'e}, Johannes and Laparra, Valero and Simoncelli, Eero P},
  booktitle={5th International Conference on Learning Representations, ICLR 2017},
  year={2017}
}

@inproceedings{he2021checkerboard,
  title={Checkerboard context model for efficient learned image compression},
  author={He, Dailan and Zheng, Yaoyan and Sun, Baocheng and Wang, Yan and Qin, Hongwei},
  booktitle={Proceedings of the IEEE/CVF Conference on Computer Vision and Pattern Recognition},
  pages={14771--14780},
  year={2021}
}

@inproceedings{li2022hybrid,
  title={Hybrid spatial-temporal entropy modelling for neural video compression},
  author={Li, Jiahao and Li, Bin and Lu, Yan},
  booktitle={Proceedings of the 30th ACM International Conference on Multimedia},
  pages={1503--1511},
  year={2022}
}

@article{chen2017targeted,
  title={Targeted backdoor attacks on deep learning systems using data poisoning},
  author={Chen, Xinyun and Liu, Chang and Li, Bo and Lu, Kimberly and Song, Dawn},
  journal={arXiv preprint arXiv:1712.05526},
  year={2017}
}

@article{gu2017badnets,
  title={Badnets: Identifying vulnerabilities in the machine learning model supply chain},
  author={Gu, Tianyu and Dolan-Gavitt, Brendan and Garg, Siddharth},
  journal={arXiv preprint arXiv:1708.06733},
  year={2017}
}

@inproceedings{liu2018trojaning,
  title={Trojaning attack on neural networks},
  author={Liu, Yingqi and Ma, Shiqing and Aafer, Yousra and Lee, Wen-Chuan and Zhai, Juan and Wang, Weihang and Zhang, Xiangyu},
  booktitle={25th Annual Network And Distributed System Security Symposium (NDSS 2018)},
  year={2018},
  organization={Internet Soc}
}

@inproceedings{liu2020reflection,
  title={Reflection backdoor: A natural backdoor attack on deep neural networks},
  author={Liu, Yunfei and Ma, Xingjun and Bailey, James and Lu, Feng},
  booktitle={Computer Vision--ECCV 2020: 16th European Conference, Glasgow, UK, August 23--28, 2020, Proceedings, Part X 16},
  pages={182--199},
  year={2020},
  organization={Springer}
}

@inproceedings{cheng2021deep,
  title={Deep feature space trojan attack of neural networks by controlled detoxification},
  author={Cheng, Siyuan and Liu, Yingqi and Ma, Shiqing and Zhang, Xiangyu},
  booktitle={Proceedings of the AAAI Conference on Artificial Intelligence},
  volume={35},
  number={2},
  pages={1148--1156},
  year={2021}
}

@inproceedings{wanet,
  title={WaNet-Imperceptible Warping-based Backdoor Attack},
  author={Nguyen, Tuan Anh and Tran, Anh Tuan},
  booktitle={International Conference on Learning Representations}
}

@inproceedings{bppattack,
  title={Bppattack: Stealthy and efficient trojan attacks against deep neural networks via image quantization and contrastive adversarial learning},
  author={Wang, Zhenting and Zhai, Juan and Ma, Shiqing},
  booktitle={Proceedings of the IEEE/CVF Conference on Computer Vision and Pattern Recognition},
  pages={15074--15084},
  year={2022}
}

@inproceedings{Ftrojan,
  title={An invisible black-box backdoor attack through frequency domain},
  author={Wang, Tong and Yao, Yuan and Xu, Feng and An, Shengwei and Tong, Hanghang and Wang, Ting},
  booktitle={European Conference on Computer Vision},
  pages={396--413},
  year={2022},
  organization={Springer}
}

@inproceedings{doan2021lira,
  title={Lira: Learnable, imperceptible and robust backdoor attacks},
  author={Doan, Khoa and Lao, Yingjie and Zhao, Weijie and Li, Ping},
  booktitle={Proceedings of the IEEE/CVF international conference on computer vision},
  pages={11966--11976},
  year={2021}
}

@inproceedings{duan2024conditional,
  title={Conditional Backdoor Attack via JPEG Compression},
  author={Duan, Qiuyu and Hua, Zhongyun and Liao, Qing and Zhang, Yushu and Zhang, Leo Yu},
  booktitle={Proceedings of the AAAI Conference on Artificial Intelligence},
  volume={38},
  number={18},
  pages={19823--19831},
  year={2024}
}

@inproceedings{yu2023backdoor,
  title={Backdoor attacks against deep image compression via adaptive frequency trigger},
  author={Yu, Yi and Wang, Yufei and Yang, Wenhan and Lu, Shijian and Tan, Yap-Peng and Kot, Alex C},
  booktitle={Proceedings of the IEEE/CVF Conference on Computer Vision and Pattern Recognition},
  pages={12250--12259},
  year={2023}
}

@article{yu2024robust,
  title={Robust and transferable backdoor attacks against deep image compression with selective frequency prior},
  author={Yu, Yi and Wang, Yufei and Yang, Wenhan and Guo, Lanqing and Lu, Shijian and Duan, Ling-Yu and Tan, Yap-Peng and Kot, Alex C},
  journal={IEEE Transactions on Pattern Analysis and Machine Intelligence},
  volume={47},
  number={3},
  pages={1674--1693},
  year={2024},
  publisher={IEEE}
}

@article{zeng2025mambaic,
  title={MambaIC: State Space Models for High-Performance Learned Image Compression},
  author={Zeng, Fanhu and Tang, Hao and Shao, Yihua and Chen, Siyu and Shao, Ling and Wang, Yan},
  journal={arXiv preprint arXiv:2503.12461},
  year={2025}
}

@article{iccv25HPCM,
  title={Learned image compression with hierarchical progressive context modeling},
  author={Li, Yuqi and Zhang, Haotian and Li, Li and Liu, Dong},
  journal={arXiv preprint arXiv:2507.19125},
  year={2025}
}

@inproceedings{wang2018recovering_sft,
  title={Recovering realistic texture in image super-resolution by deep spatial feature transform},
  author={Wang, Xintao and Yu, Ke and Dong, Chao and Loy, Chen Change},
  booktitle={Proceedings of the IEEE conference on computer vision and pattern recognition},
  pages={606--615},
  year={2018}
}

@inproceedings{zeng2021rethinking,
  title={Rethinking the backdoor attacks' triggers: A frequency perspective},
  author={Zeng, Yi and Park, Won and Mao, Z Morley and Jia, Ruoxi},
  booktitle={Proceedings of the IEEE/CVF international conference on computer vision},
  pages={16473--16481},
  year={2021}
}

@inproceedings{iclrrethinking,
  title={Rethinking CNN’s Generalization to Backdoor Attack from Frequency Domain},
  author={Rao, Quanrui and Wang, Lin and Liu, Wuying},
  booktitle={The Twelfth International Conference on Learning Representations},
  year={2024}
}

@inproceedings{li2021invisible,
  title={Invisible backdoor attack with sample-specific triggers},
  author={Li, Yuezun and Li, Yiming and Wu, Baoyuan and Li, Longkang and He, Ran and Lyu, Siwei},
  booktitle={Proceedings of the IEEE/CVF international conference on computer vision},
  pages={16463--16472},
  year={2021}
}

@article{stallkamp2012man,
  title={Man vs. computer: Benchmarking machine learning algorithms for traffic sign recognition},
  author={Stallkamp, Johannes and Schlipsing, Marc and Salmen, Jan and Igel, Christian},
  journal={Neural networks},
  volume={32},
  pages={323--332},
  year={2012},
  publisher={Elsevier}
}

@article{krizhevsky2009learning,
  title={Learning multiple layers of features from tiny images},
  author={Krizhevsky, Alex and Hinton, Geoffrey and others},
  year={2009},
  publisher={Toronto, ON, Canada}
}

@inproceedings{liu2015deep,
  title={Deep learning face attributes in the wild},
  author={Liu, Ziwei and Luo, Ping and Wang, Xiaogang and Tang, Xiaoou},
  booktitle={Proceedings of the IEEE international conference on computer vision},
  pages={3730--3738},
  year={2015}
}

@article{veldanda2020nnoculation,
  title={Nnoculation: broad spectrum and targeted treatment of backdoored dnns},
  author={Veldanda, Akshaj Kumar and Liu, Kang and Tan, Benjamin and Krishnamurthy, Prashanth and Khorrami, Farshad and Karri, Ramesh and Dolan-Gavitt, Brendan and Garg, Siddharth},
  journal={arXiv preprint arXiv:2002.08313},
  volume={3},
  pages={18},
  year={2020}
}

@inproceedings{he2016identity,
  title={Identity mappings in deep residual networks},
  author={He, Kaiming and Zhang, Xiangyu and Ren, Shaoqing and Sun, Jian},
  booktitle={Computer Vision--ECCV 2016: 14th European Conference, Amsterdam, The Netherlands, October 11--14, 2016, Proceedings, Part IV 14},
  pages={630--645},
  year={2016},
  organization={Springer}
}

@inproceedings{sandler2018mobilenetv2,
  title={Mobilenetv2: Inverted residuals and linear bottlenecks},
  author={Sandler, Mark and Howard, Andrew and Zhu, Menglong and Zhmoginov, Andrey and Chen, Liang-Chieh},
  booktitle={Proceedings of the IEEE conference on computer vision and pattern recognition},
  pages={4510--4520},
  year={2018}
}

@inproceedings{he2016deep,
  title={Deep residual learning for image recognition},
  author={He, Kaiming and Zhang, Xiangyu and Ren, Shaoqing and Sun, Jian},
  booktitle={Proceedings of the IEEE conference on computer vision and pattern recognition},
  pages={770--778},
  year={2016}
}

@inproceedings{LQ1,
  title={Discrete point-wise attack is not enough: Generalized manifold adversarial attack for face recognition},
  author={Li, Qian and Hu, Yuxiao and Liu, Ye and Zhang, Dongxiao and Jin, Xin and Chen, Yuntian},
  booktitle={Proceedings of the IEEE/CVF Conference on Computer Vision and Pattern Recognition},
  pages={20575--20584},
  year={2023}
}

@inproceedings{LQ2,
  title={Focus on Hiders: Exploring Hidden Threats for Enhancing Adversarial Training},
  author={Li, Qian and Hu, Yuxiao and Dong, Yinpeng and Zhang, Dongxiao and Chen, Yuntian},
  booktitle={Proceedings of the IEEE/CVF Conference on Computer Vision and Pattern Recognition},
  pages={24442--24451},
  year={2024}
}

@inproceedings{li2023embarrassingly,
  title={An embarrassingly simple backdoor attack on self-supervised learning},
  author={Li, Changjiang and Pang, Ren and Xi, Zhaohan and Du, Tianyu and Ji, Shouling and Yao, Yuan and Wang, Ting},
  booktitle={Proceedings of the IEEE/CVF International Conference on Computer Vision},
  pages={4367--4378},
  year={2023}
}

@inproceedings{hu2018squeeze,
  title={Squeeze-and-excitation networks},
  author={Hu, Jie and Shen, Li and Sun, Gang},
  booktitle={Proceedings of the IEEE conference on computer vision and pattern recognition},
  pages={7132--7141},
  year={2018}
}

@inproceedings{zhang2023neural,
  title={Neural Rate Control for Learned Video Compression},
  author={Zhang, Yiwei and Lu, Guo and Chen, Yunuo and Wang, Shen and Shi, Yibo and Wang, Jing and Song, Li},
  booktitle={The Twelfth International Conference on Learning Representations},
  year={2023}
}

@inproceedings{carlini2017towards,
  title={Towards evaluating the robustness of neural networks},
  author={Carlini, Nicholas and Wagner, David},
  booktitle={2017 ieee symposium on security and privacy (sp)},
  pages={39--57},
  year={2017},
  organization={Ieee}
}

@inproceedings{cohen2019certified,
  title={Certified adversarial robustness via randomized smoothing},
  author={Cohen, Jeremy and Rosenfeld, Elan and Kolter, Zico},
  booktitle={international conference on machine learning},
  pages={1310--1320},
  year={2019},
  organization={PMLR}
}

@inproceedings{costales2020live,
  title={Live trojan attacks on deep neural networks},
  author={Costales, Robby and Mao, Chengzhi and Norwitz, Raphael and Kim, Bryan and Yang, Junfeng},
  booktitle={Proceedings of the IEEE/CVF Conference on Computer Vision and Pattern Recognition Workshops},
  pages={796--797},
  year={2020}
}

@book{cox2007digital,
  title={Digital watermarking and steganography},
  author={Cox, Ingemar and Miller, Matthew and Bloom, Jeffrey and Fridrich, Jessica and Kalker, Ton},
  year={2007},
  publisher={Morgan kaufmann}
}

@inproceedings{doan2020februus,
  title={Februus: Input purification defense against trojan attacks on deep neural network systems},
  author={Doan, Bao Gia and Abbasnejad, Ehsan and Ranasinghe, Damith C},
  booktitle={Proceedings of the 36th Annual Computer Security Applications Conference},
  pages={897--912},
  year={2020}
}

@inproceedings{ganju2018property,
  title={Property inference attacks on fully connected neural networks using permutation invariant representations},
  author={Ganju, Karan and Wang, Qi and Yang, Wei and Gunter, Carl A and Borisov, Nikita},
  booktitle={Proceedings of the 2018 ACM SIGSAC conference on computer and communications security},
  pages={619--633},
  year={2018}
}

@inproceedings{gao2019strip,
  title={Strip: A defence against trojan attacks on deep neural networks},
  author={Gao, Yansong and Xu, Change and Wang, Derui and Chen, Shiping and Ranasinghe, Damith C and Nepal, Surya},
  booktitle={Proceedings of the 35th annual computer security applications conference},
  pages={113--125},
  year={2019}
}

@article{li2020invisible,
  title={Invisible backdoor attacks on deep neural networks via steganography and regularization},
  author={Li, Shaofeng and Xue, Minhui and Zhao, Benjamin Zi Hao and Zhu, Haojin and Zhang, Xinpeng},
  journal={IEEE Transactions on Dependable and Secure Computing},
  volume={18},
  number={5},
  pages={2088--2105},
  year={2020},
  publisher={IEEE}
}

@inproceedings{lin2020composite,
  title={Composite backdoor attack for deep neural network by mixing existing benign features},
  author={Lin, Junyu and Xu, Lei and Liu, Yingqi and Zhang, Xiangyu},
  booktitle={Proceedings of the 2020 ACM SIGSAC Conference on Computer and Communications Security},
  pages={113--131},
  year={2020}
}

@article{nguyen2020input,
  title={Input-aware dynamic backdoor attack},
  author={Nguyen, Tuan Anh and Tran, Anh},
  journal={Advances in Neural Information Processing Systems},
  volume={33},
  pages={3454--3464},
  year={2020}
}

@inproceedings{saha2020hidden,
  title={Hidden trigger backdoor attacks},
  author={Saha, Aniruddha and Subramanya, Akshayvarun and Pirsiavash, Hamed},
  booktitle={Proceedings of the AAAI conference on artificial intelligence},
  volume={34},
  number={07},
  pages={11957--11965},
  year={2020}
}

@article{salem2022dynamic,
  title={Dynamic Backdoor Attacks Against Machine Learning Models},
  author={Salem, Ahmed and Wen, Rui and Backes, Michael and Ma, Shiqing and Zhang, Yang},
  journal={EuroS\&P 2022},
  year={2022}
}

@inproceedings{strip,
  title={Strip: A defence against trojan attacks on deep neural networks},
  author={Gao, Yansong and Xu, Change and Wang, Derui and Chen, Shiping and Ranasinghe, Damith C and Nepal, Surya},
  booktitle={Proceedings of the 35th annual computer security applications conference},
  pages={113--125},
  year={2019}
}

@inproceedings{fine,
  title={Fine-pruning: Defending against backdooring attacks on deep neural networks},
  author={Liu, Kang and Dolan-Gavitt, Brendan and Garg, Siddharth},
  booktitle={International symposium on research in attacks, intrusions, and defenses},
  pages={273--294},
  year={2018},
  organization={Springer}
}

@article{sleeper,
  title={Sleeper agent: Scalable hidden trigger backdoors for neural networks trained from scratch},
  author={Souri, Hossein and Fowl, Liam and Chellappa, Rama and Goldblum, Micah and Goldstein, Tom},
  journal={Advances in Neural Information Processing Systems},
  volume={35},
  pages={19165--19178},
  year={2022}
}

@inproceedings{narcissus,
  title={Narcissus: A practical clean-label backdoor attack with limited information},
  author={Zeng, Yi and Pan, Minzhou and Just, Hoang Anh and Lyu, Lingjuan and Qiu, Meikang and Jia, Ruoxi},
  booktitle={Proceedings of the 2023 ACM SIGSAC Conference on Computer and Communications Security},
  pages={771--785},
  year={2023}
}

@inproceedings{issba,
  title={Invisible backdoor attack with sample-specific triggers},
  author={Li, Yuezun and Li, Yiming and Wu, Baoyuan and Li, Longkang and He, Ran and Lyu, Siwei},
  booktitle={Proceedings of the IEEE/CVF international conference on computer vision},
  pages={16463--16472},
  year={2021}
}

@article{jamil2023learning,
  title={Learning-driven lossy image compression: A comprehensive survey},
  author={Jamil, Sonain and Piran, Md Jalil and Rahman, MuhibUr and Kwon, Oh-Jin},
  journal={Engineering Applications of Artificial Intelligence},
  volume={123},
  pages={106361},
  year={2023},
  publisher={Elsevier}
}

@article{hu2021learning,
  title={Learning end-to-end lossy image compression: A benchmark},
  author={Hu, Yueyu and Yang, Wenhan and Ma, Zhan and Liu, Jiaying},
  journal={IEEE Transactions on Pattern Analysis and Machine Intelligence},
  volume={44},
  number={8},
  pages={4194--4211},
  year={2021},
  publisher={IEEE}
}

@article{yang2023lossy,
  title={Lossy image compression with conditional diffusion models},
  author={Yang, Ruihan and Mandt, Stephan},
  journal={Advances in Neural Information Processing Systems},
  volume={36},
  pages={64971--64995},
  year={2023}
}

@inproceedings{song2021variable,
  title={Variable-rate deep image compression through spatially-adaptive feature transform},
  author={Song, Myungseo and Choi, Jinyoung and Han, Bohyung},
  booktitle={Proceedings of the IEEE/CVF international conference on computer vision},
  pages={2380--2389},
  year={2021}
}

@article{roi1,
  title={End-to-end optimized ROI image compression},
  author={Cai, Chunlei and Chen, Li and Zhang, Xiaoyun and Gao, Zhiyong},
  journal={IEEE Transactions on Image Processing},
  volume={29},
  pages={3442--3457},
  year={2019},
  publisher={IEEE}
}

@inproceedings{roi2,
  title={Variable rate roi image compression optimized for visual quality},
  author={Ma, Yi and Zhai, Yongqi and Yang, Chunhui and Yang, Jiayu and Wang, Ruofan and Zhou, Jing and Li, Kai and Chen, Ying and Wang, Ronggang},
  booktitle={Proceedings of the IEEE/CVF Conference on Computer Vision and Pattern Recognition},
  pages={1936--1940},
  year={2021}
}

@inproceedings{roi3,
  title={End-to-End Learned ROI Image Compression.},
  author={Akutsu, Hiroaki and Naruko, Takahiro},
  booktitle={CVPR Workshops},
  pages={0},
  year={2019}
}

@article{roi5,
  title={Region of interest coding in JPEG 2000},
  author={Askel{\"o}f, Joel and Carlander, Mathias Larsson and Christopoulos, Charilaos},
  journal={Signal Processing: Image Communication},
  volume={17},
  number={1},
  pages={105--111},
  year={2002},
  publisher={Elsevier}
}

@inproceedings{roi6,
  title={JPEG 2000 and region of interest coding},
  author={Bradley, Andrew P and Stentiford, Fred WM},
  booktitle={Digital Image Computing Techniques and Applications},
  volume={2},
  pages={1--6},
  year={2002},
  organization={Citeseer}
}

@article{jpeg2000,
  title={The JPEG 2000 still image compression standard},
  author={Skodras, Athanassios and Christopoulos, Charilaos and Ebrahimi, Touradj},
  journal={IEEE Signal processing magazine},
  volume={18},
  number={5},
  pages={36--58},
  year={2001},
  publisher={IEEE}
}

@article{unoc,
  title={S2CFormer: Reorienting Learned Image Compression from Spatial Interaction to Channel Aggregation},
  author={Chen, Yunuo and Li, Qian and He, Bing and Feng, Donghui and Wu, Ronghua and Wang, Qi and Song, Li and Lu, Guo and Zhang, Wenjun},
  journal={arXiv preprint arXiv:2502.00700},
  year={2025}
}

@article{chen2026nefic,
  title={Next-Frame Decoding for Ultra-Low-Bitrate Image Compression with Video Diffusion Priors},
  author={Chen, Yunuo and Zhou, Chuqin and Li, Jiangchuan and Ling, Xiaoyue and He, Bing and Dai, Jincheng and Song, Li and Lu, Guo},
  journal={arXiv preprint arXiv:2603.15129},
  year={2026},
  url={https://arxiv.org/abs/2603.15129}
}

@inproceedings{saha,
  title={Hidden trigger backdoor attacks},
  author={Saha, Aniruddha and Subramanya, Akshayvarun and Pirsiavash, Hamed},
  booktitle={Proceedings of the AAAI conference on artificial intelligence},
  volume={34},
  number={07},
  pages={11957--11965},
  year={2020}
}

@article{turner,
  title={Label-consistent backdoor attacks},
  author={Turner, Alexander and Tsipras, Dimitris and Madry, Aleksander},
  journal={arXiv preprint arXiv:1912.02771},
  year={2019}
}

@inproceedings{ROIJPEG,
author = {Bařina, David and Klíma, Ondřej},
year = {2022},
month = {01},
pages = {1-5},
title = {Region of interest in JPEG},
doi = {10.24132/CSRN.3201.1}
}

@inproceedings{ROIBPG,
  title={Medical image compression based on region of interest using better portable graphics (BPG)},
  author={Yee, David and Soltaninejad, Sara and Hazarika, Deborsi and Mbuyi, Gaylord and Barnwal, Rishi and Basu, Anup},
  booktitle={2017 IEEE international conference on systems, man, and cybernetics (SMC)},
  pages={216--221},
  year={2017},
  organization={IEEE}
}

@article{zhang2025sparse,
  title={A sparse and invisible targeted backdoor attack in federated learning},
  author={Zhang, Qikun and Yu, Mengyang and Wang, Ruifang and Li, Yongjiao and Yuan, Junling and Tan, Yu-an},
  journal={Journal of King Saud University Computer and Information Sciences},
  volume={37},
  number={6},
  pages={134},
  year={2025},
  publisher={Springer}
}

@article{yang2024sampdetox,
  title={Sampdetox: Black-box backdoor defense via perturbation-based sample detoxification},
  author={Yang, Yanxin and Jia, Chentao and Yan, DengKe and Hu, Ming and Li, Tianlin and Xie, Xiaofei and Wei, Xian and Chen, Mingsong},
  journal={Advances in Neural Information Processing Systems},
  volume={37},
  pages={121236--121264},
  year={2024}
}

@inproceedings{
shi2023black_zip_nips23,
title={Black-box Backdoor Defense via Zero-shot Image Purification},
author={Shi, Yucheng and Du, Mengnan and Wu, Xuansheng and Guan, Zihan and Sun, Jin and Liu, Ninghao},
booktitle={Thirty-seventh Conference on Neural Information Processing Systems},
year={2023},
}

@article{min2023towards_FST,
  title={Towards stable backdoor purification through feature shift tuning},
  author={Min, Rui and Qin, Zeyu and Shen, Li and Cheng, Minhao},
  journal={Advances in Neural Information Processing Systems},
  volume={36},
  pages={75286--75306},
  year={2023}
}

@inproceedings{
chen2026contentaware,
title={Content-Aware Mamba for Learned Image Compression},
author={Yunuo Chen and Zezheng Lyu and Bing He and Hongwei Hu and Qi Wang and Yuan Tian and Li Song and Wenjun Zhang and Guo Lu},
booktitle={The Fourteenth International Conference on Learning Representations},
year={2026},
url={https://openreview.net/forum?id=WwDNiisZQm}
}

@article{chen2026adaptive,
  title={Adaptive Learned Image Compression with Graph Neural Networks},
  author={Chen, Yunuo and He, Bing and Lyu, Zezheng and Hu, Hongwei and Gu, Qunshan and Tian, Yuan and Lu, Guo},
  journal={arXiv preprint arXiv:2603.25316},
  year={2026}
}

@inproceedings{li2025differentiable,
  title={Differentiable vmaf: A trainable metric for optimizing video compression codec},
  author={Li, Jiangchuan and Zhou, Chuqin and Chen, Yunuo and Lu, Guo},
  booktitle={2025 IEEE International Symposium on Circuits and Systems (ISCAS)},
  pages={1--5},
  year={2025},
  organization={IEEE}
}

@article{wang5268955gaba,
  title={Gaba: A General Anti-Compression Backdoor Attack Using the Characteristic of Image Compression},
  author={Wang, Wenjie and Cong, Lin and Chen, Honglong and Han, Lifeng and Gao, Yudong and Liu, Xiaolong},
  journal={Available at SSRN 5268955}
}

@article{xue2023compression-resist,
  title={Compression-resistant backdoor attack against deep neural networks},
  author={Xue, Mingfu and Wang, Xin and Sun, Shichang and Zhang, Yushu and Wang, Jian and Liu, Weiqiang},
  journal={Applied Intelligence},
  volume={53},
  number={17},
  pages={20402--20417},
  year={2023},
  publisher={Springer}
}

@article{yang2023everyone,
  title={Everyone Can Attack: Repurpose Lossy Compression as a Natural Backdoor Attack},
  author={Yang, Sze Jue and Nguyen, Quang and Chan, Chee Seng and Doan, Khoa D},
  journal={arXiv preprint arXiv:2308.16684},
  year={2023}
}

@inproceedings{huynh2024combat_add1,
  title={Combat: Alternated training for effective clean-label backdoor attacks},
  author={Huynh, Tran and Nguyen, Dang and Pham, Tung and Tran, Anh},
  booktitle={Proceedings of the AAAI Conference on Artificial Intelligence},
  volume={38},
  number={3},
  pages={2436--2444},
  year={2024}
}

@inproceedings{jiang2023color_add2,
  title={Color backdoor: A robust poisoning attack in color space},
  author={Jiang, Wenbo and Li, Hongwei and Xu, Guowen and Zhang, Tianwei},
  booktitle={Proceedings of the IEEE/CVF conference on computer vision and pattern recognition},
  pages={8133--8142},
  year={2023}
}

@inproceedings{shin2017jpeg,
  title={{JPEG}-resistant Adversarial Images},
  author={Shin, Richard and Song, Dawn},
  booktitle={Machine Learning and Computer Security Workshop},
  year={2017},
  url={https://machine-learning-and-security.github.io/papers/mlsec17_paper_54.pdf}
}

@inproceedings{shi2021jpeg,
  title={On Generating {JPEG} Adversarial Images},
  author={Shi, Mengte and Li, Sheng and Yin, Zhaoxia and Zhang, Xinpeng and Qian, Zhenxing},
  booktitle={2021 IEEE International Conference on Multimedia and Expo (ICME)},
  pages={1--6},
  year={2021},
  doi={10.1109/ICME51207.2021.9428243}
}

@article{bonnet2022quantized,
  title={Generating Adversarial Images in Quantized Domains},
  author={Bonnet, Benoit and Furon, Teddy and Bas, Patrick},
  journal={IEEE Transactions on Information Forensics and Security},
  volume={17},
  pages={373--385},
  year={2022},
  doi={10.1109/TIFS.2021.3138616}
}

@inproceedings{athalye2018synthesizing,
  title={Synthesizing Robust Adversarial Examples},
  author={Athalye, Anish and Engstrom, Logan and Ilyas, Andrew and Kwok, Kevin},
  booktitle={Proceedings of the 35th International Conference on Machine Learning},
  volume={80},
  pages={284--293},
  year={2018},
  publisher={PMLR},
  url={https://proceedings.mlr.press/v80/athalye18b.html}
}

@inproceedings{guo2020low,
  title={Low Frequency Adversarial Perturbation},
  author={Guo, Chuan and Frank, Jared S. and Weinberger, Kilian Q.},
  booktitle={Proceedings of The 35th Uncertainty in Artificial Intelligence Conference},
  volume={115},
  pages={1127--1137},
  year={2020},
  publisher={PMLR},
  url={https://proceedings.mlr.press/v115/guo20a.html}
}

\clearpage
\newpage
\appendix

\section{More results of all-to-one attacks on different backbones}
\label{more results}
We conduct additional experiments on all-to-one attacks using two additional backbones~\cite{he2016identity, hu2018squeeze}. The results are shown in Tab.\ref{different backbones_more}. These experiments further demonstrate the robustness and generalizability of our claims and methods.
\begin{table*}[h]
\centering
\setlength{\tabcolsep}{1.5pt}
\fontsize{6.8}{10}\selectfont
\begin{tabular}{ccccc>{\columncolor{gray!16}}cc>{\columncolor{gray!16}}cc>{\columncolor{gray!16}}cc}
\noalign{\hrule height 1pt}
\textbf{Backbone $\downarrow$} & \textbf{Dataset $\downarrow$} & \textbf{Method $\rightarrow$} & \textbf{Clean} & \textbf{FTrojan} & \cellcolor{gray!16} \textbf{FTrojan}$_{\textbf{ReA}}$ & \textbf{WaNet} & \cellcolor{gray!16} $\textbf{WaNet}_\textbf{ReA}$ & \textbf{BppAttack} & \cellcolor{gray!16} $\textbf{BppAttack}_\textbf{ReA}$ & \textbf{\quad CAA\quad} \\ \hline
\multirow{6}{*}{\shortstack{Pre-activation \\ ResNet18}} & \multirow{2}{*}{CIFAR-10} & BA(\%) & 92.93 & 92.46 & 91.35(\textcolor{teal}{\hspace{0.715em}-1.11$\downarrow$}) & 86.30 & 91.30 (\textcolor{red}{\hspace{0.715em}+5.00$\uparrow$}) & 90.61 & 90.59 (\textcolor{teal}{\hspace{0.715em}-0.02$\downarrow$}) & \textbf{92.86}  \\ \cline{3-11}
 &  & ASR(\%) & - & 11.05 & 94.19(\textcolor{red}{+83.14$\uparrow$}) & 51.25 & 98.17 (\textcolor{red}{+46.92$\uparrow$}) & 16.17 & 95.03 (\textcolor{red}{+78.86$\uparrow$}) & \textbf{99.99}  \\ \cline{2-11}
 & \multirow{2}{*}{GTSRB} & BA(\%) & 99.13 & 94.25 & 97.02(\textcolor{red}{\hspace{0.715em}+2.77$\uparrow$}) & 92.73 & 97.43 (\textcolor{red}{\hspace{0.715em}+4.70$\uparrow$}) & 89.85 & 94.08 (\textcolor{red}{\hspace{0.715em}+4.23$\uparrow$}) & \textbf{99.13}  \\ \cline{3-11}
 &  & ASR(\%) & - & 23.26 & 83.77(\textcolor{red}{+60.51$\uparrow$}) & 71.80 & 92.26 (\textcolor{red}{+20.46$\uparrow$}) & 36.83 & 85.02 (\textcolor{red}{+48.19$\uparrow$}) & \textbf{99.95}  \\ \cline{2-11}
 & \multirow{2}{*}{CelebA} & BA(\%) & 76.78 & 74.98 & 70.34(\textcolor{teal}{\hspace{0.715em}-4.64$\downarrow$}) & 76.55 & 74.83 (\textcolor{teal}{\hspace{0.715em}-1.72$\downarrow$}) & 75.71 & 70.43 (\textcolor{teal}{\hspace{0.715em}-5.28$\downarrow$}) & \textbf{76.57}  \\ \cline{3-11}
 &  & ASR(\%) & - & 27.37 & 92.91(\textcolor{red}{+65.54$\uparrow$}) & 88.87 & 95.45 (\textcolor{red}{\hspace{0.715em}+6.58$\uparrow$}) & 31.32 & 94.78 (\textcolor{red}{+63.46$\uparrow$}) & \textbf{99.93}  \\ \hline
 \multirow{6}{*}{SENet} & \multirow{2}{*}{CIFAR-10} & BA(\%) & 92.97 & 92.31 & 90.93(\textcolor{teal}{\hspace{0.715em}-1.38$\downarrow$}) & 89.23 & 91.30 (\textcolor{red}{\hspace{0.715em}+2.07$\uparrow$}) & 90.67 & 91.15 (\textcolor{red}{\hspace{0.715em}+0.48$\uparrow$}) & \textbf{93.16}  \\ \cline{3-11}
 &  & ASR(\%) & - & 11.13 & 94.25(\textcolor{red}{+83.12$\uparrow$}) & 28.05 & 97.95 (\textcolor{red}{+69.90$\uparrow$}) & 16.14 & 96.03 (\textcolor{red}{+79.89$\uparrow$}) & \textbf{99.94}  \\ \cline{2-11}
 & \multirow{2}{*}{GTSRB} & BA(\%) & 98.75 & 94.87 & 97.17(\textcolor{red}{\hspace{0.715em}+2.30$\uparrow$}) & 95.02 & 97.69 (\textcolor{red}{\hspace{0.715em}+2.67$\uparrow$}) & 91.79 & 93.10 (\textcolor{red}{\hspace{0.715em}+1.31$\uparrow$}) & \textbf{98.69}  \\ \cline{3-11}
 &  & ASR(\%) & - & 29.68 & 87.43(\textcolor{red}{+57.75$\uparrow$}) & 63.25 & 92.70 (\textcolor{red}{+29.45$\uparrow$}) & 36.07 & 84.76 (\textcolor{red}{+48.69$\uparrow$}) & \textbf{99.89}  \\ \cline{2-11}
 & \multirow{2}{*}{CelebA} & BA(\%) & 77.10 & 74.73 & 71.71(\textcolor{teal}{\hspace{0.715em}-3.02$\downarrow$}) & 76.49 & 72.10 (\textcolor{teal}{\hspace{0.715em}-4.39$\downarrow$}) & 75.59 & 71.45 (\textcolor{teal}{\hspace{0.715em}-4.14$\downarrow$}) & \textbf{77.06}  \\ \cline{3-11}
 &  & ASR(\%) & - & 30.21 & 91.58(\textcolor{red}{+61.37$\uparrow$}) & 79.00 & 96.78 (\textcolor{red}{+17.78$\uparrow$}) & 31.88 & 94.05 (\textcolor{red}{+62.17$\uparrow$}) & \textbf{99.46}  \\
\noalign{\hrule height 1pt}
\end{tabular}
\vspace{-5pt}
\caption{Comparison of methods across multiple datasets. }
\vspace{-20pt}
\label{different backbones_more}
\end{table*}

\section{Defense methods in the main text}
In this part, we introduce the Fine-Pruning \cite{fine}, and STRIP \cite{strip} defense methods we used in the main text.

\noindent\textbf{Fine-Pruning \cite{fine}.} Fine-Pruning involves selectively pruning neurons that are dormant on clean inputs but become active when triggered by a backdoor. By removing these neurons, Fine-Pruning aims to deactivate the backdoor mechanism without significantly impairing the model's performance on legitimate tasks.

\noindent\textbf{STRIP \cite{strip}.} STRIP works by introducing random perturbations to inputs and observing the behavior of the model’s predictions. If the output remains consistently similar across perturbations, the input is likely compromised by a backdoor.

\section{Experiments with more defense methods}
We experiment with more advanced open-sourced defenses: SampDetox \cite{yang2024sampdetox}, FST \cite{min2023towards_FST}, and ZIP \cite{shi2023black_zip_nips23}. Since prior invisible attacks fail under compression, we compare against BadNet under JPEG. In Tab. 5, CAA shows stronger resistance than BadNet in all defenses. The performance on FST shows that
model fine-tuning seems to be a promising defense strategy in this setting. However, purification methods appear to have limited impact on CAA. We conjecture this is because CAA is compression-driven and doesn't introduce additional noise, making it more resistant to these defenses.

\begin{table}[h]
\centering
\caption{Additional defense results on GTSRB.}
\label{tab:defense_gtsrb}
\setlength{\tabcolsep}{1.2mm}
\begin{tabular}{lcccccc}
\toprule
\multirow{2}{*}{Attack}
& \multicolumn{2}{c}{SampDetox}
& \multicolumn{2}{c}{FST}
& \multicolumn{2}{c}{ZIP} \\
\cmidrule(lr){2-3} \cmidrule(lr){4-5} \cmidrule(lr){6-7}
& BA & ASR & BA & ASR & BA & ASR \\
\midrule
BadNet & 94.76 & 1.67  & 92.97 & 1.06  & 92.48 & 3.92 \\
CAA    & 95.28 & 80.38 & 92.83 & 65.38 & 94.69 & 88.29 \\
\bottomrule
\end{tabular}
\end{table}

\section{Deployment complexity}
Invisible attacks generally require more complex deployment than visible attacks (e.g., BadNets) to achieve both stealthiness and high ASR. For example, FTrojan involves transforming between RGB and YUV, frequency-domain manipulation, block-wise DCT.
Tab. \ref{tab:complexity_gtsrb} reports the comparisons on 2K images, showing that our ROI mask computation incurs no additional overhead compared to other sample-specific triggers and has a negligible impact on encoding efficiency.

\begin{table}[h]
\centering
\caption{Time complexity (s/sample) on GTSRB resized to 2K. CPU: Intel(R) Xeon(R) Gold 6458Q with 128 threads; GPU: a single NVIDIA A800 GPU.}
\label{tab:complexity_gtsrb}
\setlength{\tabcolsep}{1.6mm}
\begin{tabular}{cccccc}
\toprule
\multicolumn{3}{c}{Sample-Specific Computation}
& \multicolumn{3}{c}{Encoding Time} \\
\cmidrule(lr){1-3} \cmidrule(lr){4-6}
ROI Mask of ReA & FTrojan Trigger & WaNet Trigger
& Standard & ReA & CAA \\
\midrule
0.002 & 0.002 & 0.003 & 0.502 & 0.515 & 0.508 \\
\bottomrule
\end{tabular}
\end{table}
\section{Robustness against steganalysis.}
Tab. \ref{tab:stegan_gtsrb} reports L2 distances between poisoned and benign images in the SRM feature space, where values of 0–5 indicate negligible differences \cite{zhang2025sparse}. Since SRM mainly relies on high-pass filters, we attribute the stealthiness of CAA to  compression-driven trigger embedding instead of injecting SRM-sensitive high-frequency artifacts.

\begin{table}[h]
\centering
\caption{Comparison of SRM feature distances on GTSRB.}
\label{tab:stegan_gtsrb}
\setlength{\tabcolsep}{3.0mm}
\begin{tabular}{ccccc}
\toprule
 & FTrojan & WaNet & CAA (2$\times$2 Mask) & CAA (8$\times$8 Mask) \\
\midrule
SRM ($\downarrow$) & 0.995 & 0.628 & 0.314 & 0.207 \\
\bottomrule
\end{tabular}
\end{table}

\section{More results on different codecs}
\label{more codecs}
Tab.~\ref{jpeg}, Tab.~\ref{jpeg2000}, Tab.~\ref{bpg}, Tab.~\ref{cheng} and Tab.~\ref{elic} present the complete comparison results of our CAA and the baseline methods on JPEG2000 \cite{jpeg2000}, BPG\cite{ROIBPG}, Cheng-2020 \cite{cheng2020learned}, and ELIC \cite{he2022elic}, respectively. CAA outperforms the baseline methods in terms of both benign accuracy (BA) and attack success rate (ASR). Notably, even when the baseline methods largely fail under lossy compression scenarios, CAA still maintains a high ASR, demonstrating its robustness and effectiveness in performing data poisoning attacks under compression-induced distortions.
\begin{table*}[ht]
\centering
\setlength{\tabcolsep}{2pt}
\fontsize{7}{10}\selectfont
\begin{tabular}{cccccccc}
\noalign{\hrule height 1pt}
\textbf{Backbone $\downarrow$} & \textbf{Dataset $\downarrow$} & \textbf{Method $\rightarrow$} & \textbf{Clean} & \textbf{FTrojan} & \textbf{WaNet} & \textbf{BppAttack} & \textbf{\qquad CAA\qquad} \\ \hline
\multirow{6}{*}{ResNet18} & \multirow{2}{*}{CIFAR-10} & BA(\%) & 91.74 & 91.02 & 80.58 & 85.27  & \textbf{91.76}  \\ \cline{3-8}
&  & ASR(\%) & - & 10.82 & 73.29  & 43.92 & \textbf{97.23}  \\ \cline{2-8}
& \multirow{2}{*}{GTSRB} & BA(\%) & 98.32 & 95.23  & 94.92 & 90.61  & \textbf{98.01}  \\ \cline{3-8}
&  & ASR(\%) & - & 30.02 & 71.97 & 43.23 & \textbf{99.79}  \\ \cline{2-8}
& \multirow{2}{*}{CelebA} & BA(\%) & 75.27 & 74.29 & 75.02 & \textbf{75.32} & 75.30  \\ \cline{3-8}
&  & ASR(\%) & - & 30.23 & 77.27 & 30.86 & \textbf{99.93}  \\ \hline
\multirow{6}{*}{\shortstack{Pre-activation \\ ResNet18}} & \multirow{2}{*}{CIFAR-10} & BA(\%) & 90.92 & 91.90 & 86.18 & 91.97 & \textbf{92.76}  \\ \cline{3-8}
&  & ASR(\%) & - & 10.87 & 50.76 & 18.28 & \textbf{100.00}  \\ \cline{2-8}
& \multirow{2}{*}{GTSRB} & BA(\%) & 95.02 & 92.97 & 94.82 & 87.49 & \textbf{95.55}  \\ \cline{3-8}
&  & ASR(\%) & - & 19.87 & 73.28 & 33.29 & \textbf{99.78}  \\ \cline{2-8}
& \multirow{2}{*}{CelebA} & BA(\%) & 75.29 & 74.47 & 73.29 & 77.38 & \textbf{78.33}  \\ \cline{3-8}
&  & ASR(\%) & - & 28.97 & 86.29 & 30.29 & \textbf{99.93}  \\ \hline
\multirow{6}{*}{MobileNetV2} & \multirow{2}{*}{CIFAR-10} & BA(\%) & 92.92 & 90.76 & 91.68 & 93.49 & \textbf{93.96}  \\ \cline{3-8}
&  & ASR(\%) & - & 11.29 & 17.37 & 12.97 & \textbf{100.00}  \\ \cline{2-8}
& \multirow{2}{*}{GTSRB} & BA(\%) & 97.29 & 96.53 & 96.28 & 93.59 & \textbf{98.35}  \\ \cline{3-8}
&  & ASR(\%) & - & 14.20 & 58.57 & 24.39 & \textbf{99.75}  \\ \cline{2-8}
& \multirow{2}{*}{CelebA} & BA(\%) & 76.39 & 75.29 & 73.29 & 75.28 & \textbf{76.80}  \\ \cline{3-8}
 &  & ASR(\%) & - & 26.39 & 76.38 & 30.19 & \textbf{99.75}  \\ \hline
\multirow{6}{*}{SENet} & \multirow{2}{*}{CIFAR-10} & BA(\%) & 91.47 & \textbf{93.29} & 91.39 & 89.28 & 93.28  \\ \cline{3-8}
&  & ASR(\%) & - & 11.38 & 26.28 & 15.39 & \textbf{100.00}  \\ \cline{2-8}
& \multirow{2}{*}{GTSRB} & BA(\%) & 98.28 & 95.29 & 96.32 & 90.36 & \textbf{98.29}  \\ \cline{3-8}
&  & ASR(\%) & - & 27.30 & 66.49 & 34.29 & \textbf{99.94}  \\ \cline{2-8}
& \multirow{2}{*}{CelebA} & BA(\%) & 76.38 & 73.91 & 77.38 & 77.01 & \textbf{77.85}  \\ \cline{3-8}
&  & ASR(\%) & - & 31.38 & 77.49 & 35.23 & \textbf{99.27}  \\
\noalign{\hrule height 1pt}
\end{tabular}
\caption{Comparison results on JPEG~\cite{ROIJPEG}.}
\label{jpeg}
\end{table*}

\begin{table*}[ht]
\centering
\setlength{\tabcolsep}{2pt}
\fontsize{7}{10}\selectfont
\begin{tabular}{cccccccc}
\noalign{\hrule height 1pt}
\textbf{Backbone $\downarrow$} & \textbf{Dataset $\downarrow$} & \textbf{Method $\rightarrow$} & \textbf{Clean} & \textbf{FTrojan} & \textbf{WaNet} & \textbf{BppAttack} & \textbf{\qquad CAA\qquad} \\ \hline
\multirow{6}{*}{ResNet18} & \multirow{2}{*}{CIFAR-10} & BA(\%) & 92.85 & 90.23 & 83.29 & 88.09  & \textbf{92.93}  \\ \cline{3-8}
&  & ASR(\%) & - & 12.35 & 75.09  & 48.27 & \textbf{99.56}  \\ \cline{2-8}
& \multirow{2}{*}{GTSRB} & BA(\%) & 98.53 & 93.96  & \textbf{97.35} & 91.90 & 97.29  \\ \cline{3-8}
&  & ASR(\%) & - & 31.96 & 75.24 & 40.43 & \textbf{99.95}  \\ \cline{2-8}
& \multirow{2}{*}{CelebA} & BA(\%) & 76.41 & 76.63 & 74.19 & 72.75 & \textbf{77.63}  \\ \cline{3-8}
&  & ASR(\%) & - & 32.40 & 80.64 & 31.76 & \textbf{99.90}  \\ \hline
\multirow{6}{*}{\shortstack{Pre-activation \\ ResNet18}} & \multirow{2}{*}{CIFAR-10} & BA(\%) & 93.98 & 91.46 & 85.50 & 91.86 & \textbf{92.49}  \\ \cline{3-8}
&  & ASR(\%) & - & 12.07 & 55.09 & 18.97 & \textbf{100.00}  \\ \cline{2-8}
& \multirow{2}{*}{GTSRB} & BA(\%) & 99.37 & 95.92 & 92.76 & 89.54 & \textbf{99.04}  \\ \cline{3-8}
&  & ASR(\%) & - & 25.85 & 75.97 & 37.92 & \textbf{99.92}  \\ \cline{2-8}
& \multirow{2}{*}{CelebA} & BA(\%) & 75.66 & 74.97 & \textbf{76.83} & 73.92 & 76.27  \\ \cline{3-8}
&  & ASR(\%) & - & 25.86 & 87.54 & 33.44 & \textbf{100.00}  \\ \hline
\multirow{6}{*}{MobileNetV2} & \multirow{2}{*}{CIFAR-10} & BA(\%) & 94.23 & 92.30 & 90.28 & 91.65 & \textbf{92.48}  \\ \cline{3-8}
&  & ASR(\%) & - & 10.87 & 15.23 & 18.34 & \textbf{99.94}  \\ \cline{2-8}
& \multirow{2}{*}{GTSRB} & BA(\%) & 98.21 & 97.29 & 93.23 & 97.35 & \textbf{98.05}  \\ \cline{3-8}
&  & ASR(\%) & - & 15.97 & 65.42 & 22.29 & \textbf{99.65}  \\ \cline{2-8}
& \multirow{2}{*}{CelebA} & BA(\%) & 76.12 & \textbf{77.87} & 74.23 & 76.97 & 77.38  \\ \cline{3-8}
 &  & ASR(\%) & - & 30.97 & 80.35 & 33.23 & \textbf{99.64}  \\ \hline
\multirow{6}{*}{SENet} & \multirow{2}{*}{CIFAR-10} & BA(\%) & 93.25 & 92.85 & 88.43 & 90.38 & \textbf{93.18}  \\ \cline{3-8}
&  & ASR(\%) & - & 10.75 & 26.43 & 18.09 & \textbf{99.91}  \\ \cline{2-8}
& \multirow{2}{*}{GTSRB} & BA(\%) & 98.95 & 94.97 & 94.23 & 92.86 & \textbf{98.10}  \\ \cline{3-8}
&  & ASR(\%) & - & 28.35 & 68.85 & 37.97 & \textbf{100.00}  \\ \cline{2-8}
& \multirow{2}{*}{CelebA} & BA(\%) & 77.64 & 73.93 & 76.30 & 76.23 & \textbf{77.47}  \\ \cline{3-8}
&  & ASR(\%) & - & 30.86 & 80.23 & 34.97 & \textbf{99.95}  \\
\noalign{\hrule height 1pt}
\end{tabular}
\caption{Comparison results on JPEG2000~\cite{jpeg2000}.}
\label{jpeg2000}
\end{table*}

\begin{table*}[ht]
\centering
\setlength{\tabcolsep}{2pt}
\fontsize{7}{10}\selectfont
\begin{tabular}{cccccccc}
\noalign{\hrule height 1pt}
\textbf{Backbone $\downarrow$} & \textbf{Dataset $\downarrow$} & \textbf{Method $\rightarrow$} & \textbf{Clean} & \textbf{FTrojan} & \textbf{WaNet} & \textbf{BppAttack} & \textbf{\qquad CAA\qquad} \\ \hline
\multirow{6}{*}{ResNet18} & \multirow{2}{*}{CIFAR-10} & BA(\%) & 93.58 & 92.10 & 90.55 & \textbf{92.48}  & 92.39  \\ \cline{3-8}
&  & ASR(\%) & - & 13.07 & 76.97  & 44.97 & \textbf{100.00}  \\ \cline{2-8}
& \multirow{2}{*}{GTSRB} & BA(\%) & 99.26 & 95.13  & 95.32 & 91.07  & \textbf{98.28}  \\ \cline{3-8}
&  & ASR(\%) & - & 30.19 & 72.02 & 45.29 & \textbf{99.64}  \\ \cline{2-8}
& \multirow{2}{*}{CelebA} & BA(\%) & 77.81 & 74.27 & 76.95 & 73.43 & \textbf{77.27}  \\ \cline{3-8}
&  & ASR(\%) & - & 30.26 & 80.75 & 31.75 & \textbf{99.85}  \\ \hline
\multirow{6}{*}{\shortstack{Pre-activation \\ ResNet18}} & \multirow{2}{*}{CIFAR-10} & BA(\%) & 91.86 & \textbf{93.98} & 84.28 & 91.85 & 92.26  \\ \cline{3-8}
&  & ASR(\%) & - & 12.65 & 50.16 & 15.53 & \textbf{100.00}  \\ \cline{2-8}
& \multirow{2}{*}{GTSRB} & BA(\%) & 99.54 & 93.72 & 93.62 & 89.86 & \textbf{99.76}  \\ \cline{3-8}
&  & ASR(\%) & - & 21.86 & 73.75 & 34.28 & \textbf{99.93}  \\ \cline{2-8}
& \multirow{2}{*}{CelebA} & BA(\%) & 75.29 & \textbf{76.91} & 73.68 & 74.29 & 76.43  \\ \cline{3-8}
&  & ASR(\%) & - & 25.29 & 86.28 & 33.29 & \textbf{100.00}  \\ \hline
\multirow{6}{*}{MobileNetV2} & \multirow{2}{*}{CIFAR-10} & BA(\%) & 93.92 & 92.96 & 90.26 & 90.22 & \textbf{93.28}  \\ \cline{3-8}
&  & ASR(\%) & - & 10.28 & 15.97 & 17.83 & \textbf{99.97}  \\ \cline{2-8}
& \multirow{2}{*}{GTSRB} & BA(\%) & 98.82 & 95.23 & 92.87 & 95.21 & \textbf{98.23}  \\ \cline{3-8}
&  & ASR(\%) & - & 13.28 & 56.29 & 25.37 & \textbf{99.99}  \\ \cline{2-8}
& \multirow{2}{*}{CelebA} & BA(\%) & 76.29 & 75.54 & 73.27 & 76.29 & \textbf{76.91}  \\ \cline{3-8}
 &  & ASR(\%) & - & 27.28 & 74.29 & 30.75 & \textbf{99.99}  \\ \hline
\multirow{6}{*}{SENet} & \multirow{2}{*}{CIFAR-10} & BA(\%) & 93.65 & \textbf{93.65} & 88.64 & 90.92 & 93.27  \\ \cline{3-8}
&  & ASR(\%) & - & 10.73 & 29.64 & 15.28 & \textbf{100.00}  \\ \cline{2-8}
& \multirow{2}{*}{GTSRB} & BA(\%) & 98.82 & 95.28 & 95.81 & 92.76 & \textbf{97.74}  \\ \cline{3-8}
&  & ASR(\%) & - & 27.29 & 62.97 & 37.54 & \textbf{100.00}  \\ \cline{2-8}
& \multirow{2}{*}{CelebA} & BA(\%) & 75.29 & 74.29 & 73.92 & 74.28 & \textbf{75.01}  \\ \cline{3-8}
&  & ASR(\%) & - & 29.75 & 80.27 & 29.86 & \textbf{99.64}  \\
\noalign{\hrule height 1pt}
\end{tabular}
\caption{Comparison results on BPG~\cite{ROIBPG}.}
\label{bpg}
\end{table*}

\begin{table*}[ht]
\centering
\setlength{\tabcolsep}{2pt}
\fontsize{7}{10}\selectfont
\begin{tabular}{cccccccc}
\noalign{\hrule height 1pt}
\textbf{Backbone $\downarrow$} & \textbf{Dataset $\downarrow$} & \textbf{Method $\rightarrow$} & \textbf{Clean} & \textbf{FTrojan} & \textbf{WaNet} & \textbf{BppAttack} & \textbf{\qquad CAA\qquad} \\ \hline
\multirow{6}{*}{ResNet18} & \multirow{2}{*}{CIFAR-10} & BA(\%) & 93.21 & 91.34 & 84.86 & 88.25  & \textbf{93.14}  \\ \cline{3-8}
&  & ASR(\%) & - & 11.89 & 77.32  & 43.73 & \textbf{100.00}  \\ \cline{2-8}
& \multirow{2}{*}{GTSRB} & BA(\%) & 99.72 & 96.28  & 95.85 & 92.97  & \textbf{99.29}  \\ \cline{3-8}
&  & ASR(\%) & - & 32.89 & 71.97 & 41.93 & \textbf{99.99}  \\ \cline{2-8}
& \multirow{2}{*}{CelebA} & BA(\%) & 76.28 & 76.10 & 75.93 & 74.27 & \textbf{76.83}  \\ \cline{3-8}
&  & ASR(\%) & - & 30.28 & 78.74 & 33.97 & \textbf{100.00}  \\ \hline
\multirow{6}{*}{\shortstack{Pre-activation \\ ResNet18}} & \multirow{2}{*}{CIFAR-10} & BA(\%) & 91.98 & 91.65 & 86.36 & 90.20 & \textbf{92.53}  \\ \cline{3-8}
&  & ASR(\%) & - & 10.73 & 50.25 & 17.27 & \textbf{100.00}  \\ \cline{2-8}
& \multirow{2}{*}{GTSRB} & BA(\%) & 99.49 & 94.28 & 91.93 & 89.82 & \textbf{99.18}  \\ \cline{3-8}
&  & ASR(\%) & - & 22.96 & 74.29 & 35.27 & \textbf{99.98}  \\ \cline{2-8}
& \multirow{2}{*}{CelebA} & BA(\%) & 76.27 & 73.92 & 75.39 & 75.82 & \textbf{76.61}  \\ \cline{3-8}
&  & ASR(\%) & - & 28.45 & 88.42 & 30.63 & \textbf{99.90}  \\ \hline
\multirow{6}{*}{MobileNetV2} & \multirow{2}{*}{CIFAR-10} & BA(\%) & 93.74 & \textbf{92.63} & 90.17 & 90.36 & 92.19  \\ \cline{3-8}
&  & ASR(\%) & - & 10.27 & 16.64 & 15.27 & \textbf{99.94}  \\ \cline{2-8}
& \multirow{2}{*}{GTSRB} & BA(\%) & 98.92 & 97.29 & 94.85 & 96.84 & \textbf{98.16}  \\ \cline{3-8}
&  & ASR(\%) & - & 17.76 & 61.23 & 24.27 & \textbf{99.25}  \\ \cline{2-8}
& \multirow{2}{*}{CelebA} & BA(\%) & 76.27 & 76.06 & 75.18 & 75.91 & \textbf{76.28}  \\ \cline{3-8}
 &  & ASR(\%) & - & 30.82 & 77.41 & 33.28 & \textbf{99.98}  \\ \hline
\multirow{6}{*}{SENet} & \multirow{2}{*}{CIFAR-10} & BA(\%) & 93.21 & 92.64 & 89.25 & 90.19 & \textbf{93.03}  \\ \cline{3-8}
&  & ASR(\%) & - & 12.85 & 32.58 & 14.29 & \textbf{100.00}  \\ \cline{2-8}
& \multirow{2}{*}{GTSRB} & BA(\%) & 98.98 & 94.14 & 96.26 & 91.19 & \textbf{98.76}  \\ \cline{3-8}
&  & ASR(\%) & - & 29.75 & 62.75 & 36.19 & \textbf{100.00}  \\ \cline{2-8}
& \multirow{2}{*}{CelebA} & BA(\%) & 77.38 & 73.28 & 77.21 & 74.29 & \textbf{77.85}  \\ \cline{3-8}
&  & ASR(\%) & - & 29.75 & 79.16 & 33.26 & \textbf{99.14}  \\
\noalign{\hrule height 1pt}
\end{tabular}
\caption{Comparison results on Cheng-2020~\cite{cheng2020learned}.}
\label{cheng}
\end{table*}

\begin{table*}[ht]
\centering
\setlength{\tabcolsep}{2pt}
\fontsize{7}{10}\selectfont
\begin{tabular}{cccccccc}
\noalign{\hrule height 1pt}
\textbf{Backbone $\downarrow$} & \textbf{Dataset $\downarrow$} & \textbf{Method $\rightarrow$} & \textbf{Clean} & \textbf{FTrojan} & \textbf{WaNet} & \textbf{BppAttack} & \textbf{\qquad CAA\qquad} \\ \hline
\multirow{6}{*}{ResNet18} & \multirow{2}{*}{CIFAR-10} & BA(\%) & 93.79 & 92.32 & 85.22 & 87.35  & \textbf{93.53}  \\ \cline{3-8}
&  & ASR(\%) & - & 12.64 & 78.32  & 46.73 & \textbf{99.77}  \\ \cline{2-8}
& \multirow{2}{*}{GTSRB} & BA(\%) & 99.62 & 94.28 & 95.93 & 92.97 & \textbf{99.18}  \\ \cline{3-8}
&  & ASR(\%) & - & 29.75 & 73.92 & 41.86 & \textbf{99.98}  \\ \cline{2-8}
& \multirow{2}{*}{CelebA} & BA(\%) & 76.18 & 74.28 & 76.18 & 75.28 & \textbf{76.37}  \\ \cline{3-8}
&  & ASR(\%) & - & 29.64 & 80.28 & 30.36 & \textbf{98.28}  \\ \hline
\multirow{6}{*}{\shortstack{Pre-activation \\ ResNet18}} & \multirow{2}{*}{CIFAR-10} & BA(\%) & 92.73 & 91.36 & 86.28 & 91.65 & \textbf{92.46}  \\ \cline{3-8}
&  & ASR(\%) & - & 11.18 & 50.63 & 17.87 & \textbf{99.85}  \\ \cline{2-8}
& \multirow{2}{*}{GTSRB} & BA(\%) & 99.64 & 93.75 & 93.27 & 91.74 & \textbf{99.36}  \\ \cline{3-8}
&  & ASR(\%) & - & 21.85 & 73.64 & 33.92 & \textbf{100.00}  \\ \cline{2-8}
& \multirow{2}{*}{CelebA} & BA(\%) & 77.64 & 74.62 & 76.84 & 75.28 & \textbf{77.42}  \\ \cline{3-8}
&  & ASR(\%) & - & 25.28 & 85.28 & 30.95 & \textbf{100.00}  \\ \hline
\multirow{6}{*}{MobileNetV2} & \multirow{2}{*}{CIFAR-10} & BA(\%) & 92.29 & \textbf{93.86} & 90.74 & 90.01 & 92.37  \\ \cline{3-8}
&  & ASR(\%) & - & 13.28 & 14.12 & 16.86 & \textbf{99.42}  \\ \cline{2-8}
& \multirow{2}{*}{GTSRB} & BA(\%) & 98.10 & 96.35 & 93.84 & 95.01 & \textbf{98.96}  \\ \cline{3-8}
&  & ASR(\%) & - & 16.34 & 58.09 & 24.52 & \textbf{100.00}  \\ \cline{2-8}
& \multirow{2}{*}{CelebA} & BA(\%) & 78.59 & 77.74 & 75.38 & 75.87 & \textbf{78.25}  \\ \cline{3-8}
 &  & ASR(\%) & - & 30.78 & 77.77 & 30.85 & \textbf{98.32}  \\ \hline
\multirow{6}{*}{SENet} & \multirow{2}{*}{CIFAR-10} & BA(\%) & 93.82 & 92.45 & 88.95 & 90.83 & \textbf{93.92}  \\ \cline{3-8}
&  & ASR(\%) & - & 13.94 & 30.28 & 20.73 & \textbf{99.73}  \\ \cline{2-8}
& \multirow{2}{*}{GTSRB} & BA(\%) & 98.48 & 95.89 & 96.99 & 93.92 & \textbf{98.82}  \\ \cline{3-8}
&  & ASR(\%) & - & 27.32 & 63.54 & 37.32 & \textbf{98.35}  \\ \cline{2-8}
& \multirow{2}{*}{CelebA} & BA(\%) & 76.53 & 75.08 & 76.44 & 74.52 & \textbf{77.95}  \\ \cline{3-8}
&  & ASR(\%) & - & 29.74 & 81.46 & 32.56 & \textbf{100.00}  \\
\noalign{\hrule height 1pt}
\end{tabular}
\caption{Comparison results on ELIC \cite{he2022elic}.}
\label{elic}
\end{table*}

\clearpage
\section{Compression-Aware Trigger Optimization}
\label{app:adaptive}

Compression-resistant adversarial evasion provides relevant optimization strategies, including differentiable JPEG approximations~\cite{shin2017jpeg}, attacks in JPEG and quantized domains~\cite{shi2021jpeg,bonnet2022quantized}, expectation over transformations~\cite{athalye2018synthesizing}, and low-frequency perturbation design~\cite{guo2020low}. Adversarial evasion targets immediate misclassification by a fixed classifier, whereas backdoor poisoning must induce a training-time association in a subsequently trained model. We therefore adapt compression-aware optimization to a backdoor baseline rather than treating adversarial examples as direct poisoning baselines.

\noindent\textbf{Compression-Aware Narcissus.}
Starting from the original Narcissus settings~\cite{narcissus}, we insert a frozen, differentiable Cheng-2020 codec between the triggered image and the surrogate classifier during trigger optimization. Gradients propagate through the codec, but only the trigger is optimized; the codec, surrogate classifier, and victim-side training procedure remain unchanged. We evaluate the resulting trigger on GTSRB with Cheng-2020 and then transfer the same trigger to ELIC without re-optimization.

\begin{table}[htbp]
\centering
\small
\setlength{\tabcolsep}{3pt}
\begin{tabular}{lllrr}
\toprule
Method & Optimization codec & Evaluation codec & BA (\%) & ASR (\%) \\
\midrule
Vanilla Narcissus & None & Cheng-2020 & 93.71 & 23.54 \\
Compression-Aware Narcissus & Cheng-2020 & Cheng-2020 & 94.19 & 67.53 \\
Compression-Aware Narcissus & Cheng-2020 & ELIC & 90.07 & 11.07 \\
\bottomrule
\end{tabular}
\caption{Compression-aware Narcissus on GTSRB. The ELIC row reuses the trigger optimized with Cheng-2020.}
\label{tab:adaptive-narcissus}
\end{table}

Matched-codec optimization increases ASR from 23.54\% to 67.53\%, showing that ROI coding is not the only way to improve compression robustness. ASR falls to 11.07\% when this trigger is transferred to ELIC. This result demonstrates codec sensitivity for the evaluated trigger and codec pair, rather than a general failure of adaptive optimization to transfer. These experiments are separate from the Mbt-2018 reactivation comparison in Tab.~\ref{clean-label}: UAR preserves an existing trigger through ROI coding, while Compression-Aware Narcissus redesigns a trigger for a known differentiable codec.

\noindent\textbf{Interpreting the compression-based baselines.}
Tab.~\ref{compression-aware} evaluates ECA~\cite{yang2023everyone} and CBA~\cite{duan2024conditional} when both benign and poisoned samples undergo a shared compression stage. ECA uses an additional JPEG or WEBP transformation to create its poisoned samples. Shared encoding gives benign samples compression artifacts as well and modifies the artifacts already present in poisoned samples, reducing the distinction on which this trigger relies. For CBA, the comparison uses a fixed data-side JPEG transformation rather than retaining its original joint optimization of the compression operator and target classifier. The lower ASRs therefore describe transfer to the evaluated shared-compression setting, rather than performance under the baselines' original protocols.

\clearpage
\section{Additional Training and Robustness Experiments}
\label{app:training-robustness}

\subsection{Lower poisoning rate}
Each invisible baseline and its UAR variant use the same poisoning budget: 5\% for FTrojan, 10\% for WaNet, and 20\% for BppAttack, following their original configurations. The main CAA comparison uses 5\% poisoning. To evaluate a lower budget, we reduce CAA's poisoning rate to 2\% on CIFAR-10 using ResNet18 and Mbt-2018, while keeping the ROI mask, codec configuration, and victim training settings unchanged.

\begin{table}[htbp]
\centering
\small
\begin{tabular}{lrrr}
\toprule
Dataset & Poisoning rate (\%) & BA (\%) & ASR (\%) \\
\midrule
CIFAR-10 & 5 & 93.28 & 100.00 \\
CIFAR-10 & 2 & 92.10 & 98.91 \\
\bottomrule
\end{tabular}
\caption{CAA poisoning-rate ablation on CIFAR-10 with ResNet18 and Mbt-2018.}
\label{tab:poisoning-rate}
\end{table}

CAA retains 98.91\% ASR at 2\% poisoning in this configuration. This two-point comparison supports effectiveness at a lower budget on CIFAR-10, but does not determine the minimum poisoning rate or establish the same behavior across all datasets.

\subsection{Clean data outside the attacker-controlled encoding pipeline}
We use a fixed-size GTSRB experiment to isolate the effect of benign-data encoding. We progressively replace attacker-encoded benign samples with uncompressed clean samples, yielding three benign-subset compositions: all attacker-encoded, 50\% attacker-encoded and 50\% uncompressed, and all uncompressed. In the 50/50 setting, an equal number of attacker-encoded benign samples is subsampled when uncompressed samples are introduced. The poisoned subset, poisoning rate, total training-set size, and victim training configuration remain fixed. This is a controlled proxy for clean-data mixing, rather than an experiment that enlarges the training set and dilutes the poisoning rate.

\begin{table}[htbp]
\centering
\small
\setlength{\tabcolsep}{4pt}
\begin{tabular}{lccc}
\toprule
Method & All attacker-encoded & \makecell{50\% attacker-encoded\\50\% uncompressed} & All uncompressed \\
\midrule
FTrojan$_{\rm ReA}$ (UAR) & 97.43 / 89.34 & 96.87 / 88.71 & 98.69 / 90.02 \\
CAA & 99.13 / 100.00 & 99.00 / 100.00 & 99.13 / 100.00 \\
\bottomrule
\end{tabular}
\caption{BA / ASR (\%) under fixed-size benign-data replacement on GTSRB. Column headings describe the benign subset only.}
\label{tab:clean-data-mixing}
\end{table}

FTrojan$_{\rm ReA}$ achieves 88.71--90.02\% ASR and CAA achieves 100\% across these settings. Thus, the evaluated attacks remain effective when their poisoned samples are mixed with benign samples that have not undergone attacker-controlled encoding. The comparison does not measure the effect of adding a larger external dataset or introducing a distribution shift.

\subsection{Benign accuracy under JPEG recompression}
Tab.~\ref{tab:recompression-utility} supplements the ASRs in Tab.~\ref{tab:recompression_gtsrb} with their corresponding BA. JPEG recompression is applied after the initial ROI-based compression on GTSRB. The no-recompression column gives the reference for each attack and mask.

\begin{table}[htbp]
\centering
\small
\setlength{\tabcolsep}{4pt}
\begin{tabular}{lcccc}
\toprule
Attack & No recompression & Q50 & Q75 & Q90 \\
\midrule
FTrojan$_{\rm ReA}$ & 97.43 / 89.34 & 90.21 / 10.83 & 93.08 / 17.87 & 96.31 / 25.56 \\
CAA ($2\times2$) & 99.11 / 100.00 & 92.73 / 79.59 & 96.81 / 95.65 & 99.05 / 97.08 \\
CAA ($8\times8$) & 99.13 / 100.00 & 90.02 / 73.27 & 92.84 / 88.87 & 98.97 / 96.57 \\
\bottomrule
\end{tabular}
\caption{BA / ASR (\%) under an additional JPEG compression on GTSRB.}
\label{tab:recompression-utility}
\end{table}

Q50 substantially reduces FTrojan$_{\rm ReA}$ ASR and partially mitigates CAA, but also lowers BA by 6.38--9.11 percentage points relative to the corresponding references. Q90 largely preserves CAA's BA while leaving ASR above 96\%. These results show a defense--utility trade-off for uniform recompression.

\subsection{Low-pass preprocessing before victim training}
We apply the same Gaussian filter to all decompressed benign and poisoned training images before training ResNet18 on GTSRB, leaving the remaining training settings unchanged. This training-time preprocessing experiment is distinct from the test-time Gaussian perturbations reported in the main text.

\begin{table}[htbp]
\centering
\small
\begin{tabular}{lrrr}
\toprule
CAA mask & No filtering & Gaussian $3\times3$ & Gaussian $7\times7$ \\
\midrule
$2\times2$ & 100.00 & 100.00 & 96.21 \\
$8\times8$ & 100.00 & 93.87 & 80.73 \\
\bottomrule
\end{tabular}
\caption{ASR (\%) after Gaussian preprocessing of the training data on GTSRB with ResNet18.}
\label{tab:training-lowpass}
\end{table}

Stronger filtering reduces ASR, particularly for the finer $8\times8$ pattern. The coarser pattern retains higher ASR under both tested filters, consistent with the recompression comparison. The reported ASRs alone do not quantify the benign-accuracy cost of this preprocessing defense.

\subsection{ROI-mask selection and mixtures}
We compare $2\times2$ and $8\times8$ checkerboards, concentric squares, and a variant that randomly selects one of these three masks for each poisoned sample. Tab.~\ref{tab:mask-selection} reports single-compression BA/ASR and ASR after an additional JPEG encoding.

\begin{table}[htbp]
\centering
\small
\setlength{\tabcolsep}{4pt}
\begin{tabular}{lrrrr}
\toprule
Mask configuration & BA (\%) & ASR (\%) & Q75 ASR (\%) & Q90 ASR (\%) \\
\midrule
$2\times2$ checkerboard & 99.11 & 100.00 & 95.65 & 97.08 \\
$8\times8$ checkerboard & 99.13 & 100.00 & 88.87 & 96.57 \\
Concentric squares & 99.20 & 100.00 & 96.37 & 98.13 \\
Three-mask pool (random selection) & 98.46 & 100.00 & 85.32 & 95.27 \\
\bottomrule
\end{tabular}
\caption{ROI-mask selection and JPEG recompression.}
\label{tab:mask-selection}
\end{table}

All four configurations achieve 100\% ASR before recompression, but differ afterward. The three-mask pool preserves single-compression attack success while having lower recompression ASR than the individual patterns at the tested qualities. Together with Tab.~\ref{tab:fft-detection}, the checkerboard results indicate a trade-off: finer allocation is harder for the matched detector to distinguish, but more sensitive to recompression and low-pass filtering. The detection experiment does not evaluate concentric squares or the mask pool, so it does not establish a detection advantage for mask diversity.

\clearpage
\newpage
\section*{NeurIPS Paper Checklist}

\begin{enumerate}

\item {\bf Claims}
    \item[] Question: Do the main claims made in the abstract and introduction accurately reflect the paper's contributions and scope?
    \item[] Answer: \answerYes{}
    \item[] Justification: Sec.~\ref{intro}
    \item[] Guidelines:
    \begin{itemize}
        \item The answer \answerNA{} means that the abstract and introduction do not include the claims made in the paper.
        \item The abstract and/or introduction should clearly state the claims made, including the contributions made in the paper and important assumptions and limitations. A \answerNo{} or \answerNA{} answer to this question will not be perceived well by the reviewers.
        \item The claims made should match theoretical and experimental results, and reflect how much the results can be expected to generalize to other settings.
        \item It is fine to include aspirational goals as motivation as long as it is clear that these goals are not attained by the paper.
    \end{itemize}

\item {\bf Limitations}
    \item[] Question: Does the paper discuss the limitations of the work performed by the authors?
    \item[] Answer: \answerYes{}
    \item[] Justification: Sec.~\ref{conclusion}
    \item[] Guidelines:
    \begin{itemize}
        \item The answer \answerNA{} means that the paper has no limitation while the answer \answerNo{} means that the paper has limitations, but those are not discussed in the paper.
        \item The authors are encouraged to create a separate ``Limitations'' section in their paper.
        \item The paper should point out any strong assumptions and how robust the results are to violations of these assumptions (e.g., independence assumptions, noiseless settings, model well-specification, asymptotic approximations only holding locally). The authors should reflect on how these assumptions might be violated in practice and what the implications would be.
        \item The authors should reflect on the scope of the claims made, e.g., if the approach was only tested on a few datasets or with a few runs. In general, empirical results often depend on implicit assumptions, which should be articulated.
        \item The authors should reflect on the factors that influence the performance of the approach. For example, a facial recognition algorithm may perform poorly when image resolution is low or images are taken in low lighting. Or a speech-to-text system might not be used reliably to provide closed captions for online lectures because it fails to handle technical jargon.
        \item The authors should discuss the computational efficiency of the proposed algorithms and how they scale with dataset size.
        \item If applicable, the authors should discuss possible limitations of their approach to address problems of privacy and fairness.
        \item While the authors might fear that complete honesty about limitations might be used by reviewers as grounds for rejection, a worse outcome might be that reviewers discover limitations that aren't acknowledged in the paper. The authors should use their best judgment and recognize that individual actions in favor of transparency play an important role in developing norms that preserve the integrity of the community. Reviewers will be specifically instructed to not penalize honesty concerning limitations.
    \end{itemize}

\item {\bf Theory assumptions and proofs}
    \item[] Question: For each theoretical result, does the paper provide the full set of assumptions and a complete (and correct) proof?
    \item[] Answer: \answerNA{}
    \item[] Justification: the paper does not include theoretical results.
    \item[] Guidelines:
    \begin{itemize}
        \item The answer \answerNA{} means that the paper does not include theoretical results.
        \item All the theorems, formulas, and proofs in the paper should be numbered and cross-referenced.
        \item All assumptions should be clearly stated or referenced in the statement of any theorems.
        \item The proofs can either appear in the main paper or the supplemental material, but if they appear in the supplemental material, the authors are encouraged to provide a short proof sketch to provide intuition.
        \item Inversely, any informal proof provided in the core of the paper should be complemented by formal proofs provided in appendix or supplemental material.
        \item Theorems and Lemmas that the proof relies upon should be properly referenced.
    \end{itemize}

    \item {\bf Experimental result reproducibility}
    \item[] Question: Does the paper fully disclose all the information needed to reproduce the main experimental results of the paper to the extent that it affects the main claims and/or conclusions of the paper (regardless of whether the code and data are provided or not)?
    \item[] Answer: \answerYes{}
    \item[] Justification: Sec.~\ref{exps}
    \item[] Guidelines:
    \begin{itemize}
        \item The answer \answerNA{} means that the paper does not include experiments.
        \item If the paper includes experiments, a \answerNo{} answer to this question will not be perceived well by the reviewers: Making the paper reproducible is important, regardless of whether the code and data are provided or not.
        \item If the contribution is a dataset and\slash or model, the authors should describe the steps taken to make their results reproducible or verifiable.
        \item Depending on the contribution, reproducibility can be accomplished in various ways. For example, if the contribution is a novel architecture, describing the architecture fully might suffice, or if the contribution is a specific model and empirical evaluation, it may be necessary to either make it possible for others to replicate the model with the same dataset, or provide access to the model. In general. releasing code and data is often one good way to accomplish this, but reproducibility can also be provided via detailed instructions for how to replicate the results, access to a hosted model (e.g., in the case of a large language model), releasing of a model checkpoint, or other means that are appropriate to the research performed.
        \item While NeurIPS does not require releasing code, the conference does require all submissions to provide some reasonable avenue for reproducibility, which may depend on the nature of the contribution. For example
        \begin{enumerate}
            \item If the contribution is primarily a new algorithm, the paper should make it clear how to reproduce that algorithm.
            \item If the contribution is primarily a new model architecture, the paper should describe the architecture clearly and fully.
            \item If the contribution is a new model (e.g., a large language model), then there should either be a way to access this model for reproducing the results or a way to reproduce the model (e.g., with an open-source dataset or instructions for how to construct the dataset).
            \item We recognize that reproducibility may be tricky in some cases, in which case authors are welcome to describe the particular way they provide for reproducibility. In the case of closed-source models, it may be that access to the model is limited in some way (e.g., to registered users), but it should be possible for other researchers to have some path to reproducing or verifying the results.
        \end{enumerate}
    \end{itemize}

\item {\bf Open access to data and code}
    \item[] Question: Does the paper provide open access to the data and code, with sufficient instructions to faithfully reproduce the main experimental results, as described in supplemental material?
    \item[] Answer: \answerNo{}
    \item[] Justification: We plan to release the implementation.
    \item[] Guidelines:
    \begin{itemize}
        \item The answer \answerNA{} means that paper does not include experiments requiring code.
        \item Please see the NeurIPS code and data submission guidelines (\url{https://neurips.cc/public/guides/CodeSubmissionPolicy}) for more details.
        \item While we encourage the release of code and data, we understand that this might not be possible, so \answerNo{} is an acceptable answer. Papers cannot be rejected simply for not including code, unless this is central to the contribution (e.g., for a new open-source benchmark).
        \item The instructions should contain the exact command and environment needed to run to reproduce the results. See the NeurIPS code and data submission guidelines (\url{https://neurips.cc/public/guides/CodeSubmissionPolicy}) for more details.
        \item The authors should provide instructions on data access and preparation, including how to access the raw data, preprocessed data, intermediate data, and generated data, etc.
        \item The authors should provide scripts to reproduce all experimental results for the new proposed method and baselines. If only a subset of experiments are reproducible, they should state which ones are omitted from the script and why.
        \item At submission time, to preserve anonymity, the authors should release anonymized versions (if applicable).
        \item Providing as much information as possible in supplemental material (appended to the paper) is recommended, but including URLs to data and code is permitted.
    \end{itemize}

\item {\bf Experimental setting/details}
    \item[] Question: Does the paper specify all the training and test details (e.g., data splits, hyperparameters, how they were chosen, type of optimizer) necessary to understand the results?
    \item[] Answer: \answerYes{}
    \item[] Justification: Sec.~\ref{exps}
    \item[] Guidelines:
    \begin{itemize}
        \item The answer \answerNA{} means that the paper does not include experiments.
        \item The experimental setting should be presented in the core of the paper to a level of detail that is necessary to appreciate the results and make sense of them.
        \item The full details can be provided either with the code, in appendix, or as supplemental material.
    \end{itemize}

\item {\bf Experiment statistical significance}
    \item[] Question: Does the paper report error bars suitably and correctly defined or other appropriate information about the statistical significance of the experiments?
    \item[] Answer: \answerYes{}
    \item[] Justification: Sec.~\ref{exps}
    \item[] Guidelines:
    \begin{itemize}
        \item The answer \answerNA{} means that the paper does not include experiments.
        \item The authors should answer \answerYes{} if the results are accompanied by error bars, confidence intervals, or statistical significance tests, at least for the experiments that support the main claims of the paper.
        \item The factors of variability that the error bars are capturing should be clearly stated (for example, train/test split, initialization, random drawing of some parameter, or overall run with given experimental conditions).
        \item The method for calculating the error bars should be explained (closed form formula, call to a library function, bootstrap, etc.)
        \item The assumptions made should be given (e.g., Normally distributed errors).
        \item It should be clear whether the error bar is the standard deviation or the standard error of the mean.
        \item It is OK to report 1-sigma error bars, but one should state it. The authors should preferably report a 2-sigma error bar than state that they have a 96\% CI, if the hypothesis of Normality of errors is not verified.
        \item For asymmetric distributions, the authors should be careful not to show in tables or figures symmetric error bars that would yield results that are out of range (e.g., negative error rates).
        \item If error bars are reported in tables or plots, the authors should explain in the text how they were calculated and reference the corresponding figures or tables in the text.
    \end{itemize}

\item {\bf Experiments compute resources}
    \item[] Question: For each experiment, does the paper provide sufficient information on the computer resources (type of compute workers, memory, time of execution) needed to reproduce the experiments?
    \item[] Answer: \answerYes{}
    \item[] Justification: Sec.~\ref{exps}
    \item[] Guidelines:
    \begin{itemize}
        \item The answer \answerNA{} means that the paper does not include experiments.
        \item The paper should indicate the type of compute workers CPU or GPU, internal cluster, or cloud provider, including relevant memory and storage.
        \item The paper should provide the amount of compute required for each of the individual experimental runs as well as estimate the total compute.
        \item The paper should disclose whether the full research project required more compute than the experiments reported in the paper (e.g., preliminary or failed experiments that didn't make it into the paper).
    \end{itemize}

\item {\bf Code of ethics}
    \item[] Question: Does the research conducted in the paper conform, in every respect, with the NeurIPS Code of Ethics \url{https://neurips.cc/public/EthicsGuidelines}?
    \item[] Answer: \answerYes{}
    \item[] Justification:  Yes
    \item[] Guidelines:
    \begin{itemize}
        \item The answer \answerNA{} means that the authors have not reviewed the NeurIPS Code of Ethics.
        \item If the authors answer \answerNo, they should explain the special circumstances that require a deviation from the Code of Ethics.
        \item The authors should make sure to preserve anonymity (e.g., if there is a special consideration due to laws or regulations in their jurisdiction).
    \end{itemize}

\item {\bf Broader impacts}
    \item[] Question: Does the paper discuss both potential positive societal impacts and negative societal impacts of the work performed?
    \item[] Answer: \answerYes{}
    \item[] Justification: This paper primarily focuses on the technical aspects of compression-aware backdoor attacks. Section 5 discusses limitations and potential detection strategies.
    \item[] Guidelines:
    \begin{itemize}
        \item The answer \answerNA{} means that there is no societal impact of the work performed.
        \item If the authors answer \answerNA{} or \answerNo, they should explain why their work has no societal impact or why the paper does not address societal impact.
        \item Examples of negative societal impacts include potential malicious or unintended uses (e.g., disinformation, generating fake profiles, surveillance), fairness considerations (e.g., deployment of technologies that could make decisions that unfairly impact specific groups), privacy considerations, and security considerations.
        \item The conference expects that many papers will be foundational research and not tied to particular applications, let alone deployments. However, if there is a direct path to any negative applications, the authors should point it out. For example, it is legitimate to point out that an improvement in the quality of generative models could be used to generate Deepfakes for disinformation. On the other hand, it is not needed to point out that a generic algorithm for optimizing neural networks could enable people to train models that generate Deepfakes faster.
        \item The authors should consider possible harms that could arise when the technology is being used as intended and functioning correctly, harms that could arise when the technology is being used as intended but gives incorrect results, and harms following from (intentional or unintentional) misuse of the technology.
        \item If there are negative societal impacts, the authors could also discuss possible mitigation strategies (e.g., gated release of models, providing defenses in addition to attacks, mechanisms for monitoring misuse, mechanisms to monitor how a system learns from feedback over time, improving the efficiency and accessibility of ML).
    \end{itemize}

\item {\bf Safeguards}
    \item[] Question: Does the paper describe safeguards that have been put in place for responsible release of data or models that have a high risk for misuse (e.g., pre-trained language models, image generators, or scraped datasets)?
    \item[] Answer: \answerNA{}
    \item[] Justification: This work does not release pretrained backdoored models or poisoned datasets. We evaluate the proposed methods using existing benchmark datasets.
    \item[] Guidelines:
    \begin{itemize}
        \item The answer \answerNA{} means that the paper poses no such risks.
        \item Released models that have a high risk for misuse or dual-use should be released with necessary safeguards to allow for controlled use of the model, for example by requiring that users adhere to usage guidelines or restrictions to access the model or implementing safety filters.
        \item Datasets that have been scraped from the Internet could pose safety risks. The authors should describe how they avoided releasing unsafe images.
        \item We recognize that providing effective safeguards is challenging, and many papers do not require this, but we encourage authors to take this into account and make a best faith effort.
    \end{itemize}

\item {\bf Licenses for existing assets}
    \item[] Question: Are the creators or original owners of assets (e.g., code, data, models), used in the paper, properly credited and are the license and terms of use explicitly mentioned and properly respected?
    \item[] Answer: \answerYes{}
    \item[] Justification: Existing assets used in the paper have been properly credited.
    \item[] Guidelines:
    \begin{itemize}
        \item The answer \answerNA{} means that the paper does not use existing assets.
        \item The authors should cite the original paper that produced the code package or dataset.
        \item The authors should state which version of the asset is used and, if possible, include a URL.
        \item The name of the license (e.g., CC-BY 4.0) should be included for each asset.
        \item For scraped data from a particular source (e.g., website), the copyright and terms of service of that source should be provided.
        \item If assets are released, the license, copyright information, and terms of use in the package should be provided. For popular datasets, \url{paperswithcode.com/datasets} has curated licenses for some datasets. Their licensing guide can help determine the license of a dataset.
        \item For existing datasets that are re-packaged, both the original license and the license of the derived asset (if it has changed) should be provided.
        \item If this information is not available online, the authors are encouraged to reach out to the asset's creators.
    \end{itemize}

\item {\bf New assets}
    \item[] Question: Are new assets introduced in the paper well documented and is the documentation provided alongside the assets?
    \item[] Answer: \answerNA{}
    \item[] Justification: None.
    \item[] Guidelines:
    \begin{itemize}
        \item The answer \answerNA{} means that the paper does not release new assets.
        \item Researchers should communicate the details of the dataset\slash code\slash model as part of their submissions via structured templates. This includes details about training, license, limitations, etc.
        \item The paper should discuss whether and how consent was obtained from people whose asset is used.
        \item At submission time, remember to anonymize your assets (if applicable). You can either create an anonymized URL or include an anonymized zip file.
    \end{itemize}

\item {\bf Crowdsourcing and research with human subjects}
    \item[] Question: For crowdsourcing experiments and research with human subjects, does the paper include the full text of instructions given to participants and screenshots, if applicable, as well as details about compensation (if any)?
    \item[] Answer: \answerNo{}
    \item[] Justification: The human inspection study is reported in Sec.\ref{exps} and partial detailed information is not included.
    \item[] Guidelines:
    \begin{itemize}
        \item The answer \answerNA{} means that the paper does not involve crowdsourcing nor research with human subjects.
        \item Including this information in the supplemental material is fine, but if the main contribution of the paper involves human subjects, then as much detail as possible should be included in the main paper.
        \item According to the NeurIPS Code of Ethics, workers involved in data collection, curation, or other labor should be paid at least the minimum wage in the country of the data collector.
    \end{itemize}

\item {\bf Institutional review board (IRB) approvals or equivalent for research with human subjects}
    \item[] Question: Does the paper describe potential risks incurred by study participants, whether such risks were disclosed to the subjects, and whether Institutional Review Board (IRB) approvals (or an equivalent approval/review based on the requirements of your country or institution) were obtained?
    \item[] Answer: \answerNo{}
    \item[] Justification: The human inspection study involved a low-risk visual perception task. Formal IRB approval was not required.
    \item[] Guidelines:
    \begin{itemize}
        \item The answer \answerNA{} means that the paper does not involve crowdsourcing nor research with human subjects.
        \item Depending on the country in which research is conducted, IRB approval (or equivalent) may be required for any human subjects research. If you obtained IRB approval, you should clearly state this in the paper.
        \item We recognize that the procedures for this may vary significantly between institutions and locations, and we expect authors to adhere to the NeurIPS Code of Ethics and the guidelines for their institution.
        \item For initial submissions, do not include any information that would break anonymity (if applicable), such as the institution conducting the review.
    \end{itemize}

\item {\bf Declaration of LLM usage}
    \item[] Question: Does the paper describe the usage of LLMs if it is an important, original, or non-standard component of the core methods in this research? Note that if the LLM is used only for writing, editing, or formatting purposes and does \emph{not} impact the core methodology, scientific rigor, or originality of the research, declaration is not required.
    \item[] Answer: \answerNA{}
    \item[] Justification: the core method development in this research does not involve LLMs as any important, original, or non-standard components.
    \item[] Guidelines:
    \begin{itemize}
        \item The answer \answerNA{} means that the core method development in this research does not involve LLMs as any important, original, or non-standard components.
        \item Please refer to our LLM policy in the NeurIPS handbook for what should or should not be described.
    \end{itemize}

\end{enumerate}

\end{document}